\documentclass[superscriptaddress,prb]{revtex4}

\usepackage{amsmath,amssymb,graphicx}

\newcommand{\be}{\begin{equation}}
\newcommand{\ee}{\end{equation}}

\newcommand{\ba}{\begin{eqnarray}}
\newcommand{\ea}{\end{eqnarray}}

\let\a=\alpha \let\b=\beta \let\g=\gamma \let\d=\delta
\let\e=\varepsilon   
\let\l=\lambda \let\m=\mu   
\let\s=\sigma \let\t=\tau  
   
\let\D=\Delta   
   
\let\ee=\epsilon \let\r=\rho \let\th=\theta \let\io=\infty

\def\ie{{i.e. }}\def\eg{{e.g. }}

 \def\VV{{\cal V}}
 
\def\NN{{\cal N}} 
\def\RR{{\cal R}}  \def\OO{{\cal O}}

 \def\xx{{\bf x}} \def\yy{{\bf y}}

\def\AA{{{\rm A}^{^{\!\!\!\!\!\!\;\circ}}}}

\def\to{\rightarrow} \def\la{\langle} \def\ra{\rangle}

\newcommand{\beq}{\begin{equation}} \newcommand{\eeq}{\end{equation}}
 \newcommand{\wt}{\widetilde}

\begin{document}

\title{Theory of the superglass phase}

\author{Giulio Biroli}
\affiliation{Institut de Physique Th{\'e}orique,
CEA, IPhT, F-91191 Gif-sur-Yvette, France
CNRS, URA 2306, F-91191 Gif-sur-Yvette, France}
\author{Claudio Chamon}
\affiliation{Physics Department, Boston University, Boston, MA 02215, USA}
\author{Francesco Zamponi}
\affiliation{CNRS-UMR 8549, Laboratoire de Physique Th\'eorique, 
\'Ecole Normale Sup\'erieure,
24 Rue Lhomond, 75231 Paris Cedex 05, France
}

\date{\today}

\begin{abstract} 
A superglass is a phase of matter which is characterized at the same
time by superfluidity and a frozen amorphous structure.  We introduce
a model of interacting bosons in three dimensions that displays this
phase unambiguously and that can be analyzed exactly or using
controlled approximations.  Employing a mapping between quantum
Hamiltonians and classical Fokker-Planck operators, we show that the
ground state wavefunction of the quantum model is proportional to the
Boltzmann measure of classical hard spheres. This connection allows us
to obtain quantitative results on static and dynamic quantum
correlation functions.  In particular, by translating known results on
the glassy dynamics of Brownian hard spheres we work out the
properties of the superglass phase and of the quantum phase
transition between the superfluid and the superglass phase.
\end{abstract}

\maketitle

\section{Introduction}
Solids do not flow. This apparently tautological statement can be
wrong in two ways. First, solidity is in general a timescale-dependent
phenomenon. Crystal or glasses, well known solids on most
experimentally relevant timescales, do flow if one waits long enough
(see Ref.~\onlinecite{pantarei} for example). Second, at very small
temperatures, quantum solids may become superfluids as suggested
theoretically in the early seventies \cite{Andreev,Leggett,Reatto}.  A
striking property of this phase of matter is the non-classical
rotational inertia: a supersolid placed at thermal equilibrium in a
rotating container does not rotate, at least not completely, for small
angular velocity $\omega$. In particular, its angular momentum is
reduced from its classical value $I_{\rm cl}\; \omega$ by a fraction
$f_s$ which is called superfluid fraction.

The first promising experimental evidence of supersolidity was found
only four years ago by Kim and Chan \cite{Chan1,Chan2} through
measurements of a non-classical rotational momentum of inertia (NCRI)
in solid He$^4$ at low temperature. Unfortunately, the physical
mechanism underlying the observed effect is still unclear. In
particular, despite intense experimental activity, it has not been
established that the observed NCRI in solid He$^4$ is accompanied by
superflow (see, however, the very recent work reported in
Ref.~\onlinecite{superflow-solid}).

Nonetheless, some agreement has been reached by now on few basic facts
\cite{balibarreview}: theoretical works have revealed that equilibrium
He$^4$ crystals at zero temperature are commensurate and {\it not} 
supersolids \cite{Ceperley,BPS0,Vi08}, or at
least have an extremely small superfluid fraction, much too small to
account for experimental findings (see also
Ref.~\onlinecite{Pomeau}). Experiments have shown that the supersolid
critical temperature and the superfluid density depend sensitively on
the detailed preparation history of the solid samples \cite{reppy1}
and on the presence of minute fractions of He$^3$
impurities\cite{Chan1,Chan2}.  The more disordered are the He$^4$
solids the larger is the supersolid signal. It can reach surprisingly
large values when the solid is formed by a rapid freezing from the
normal phase. It can be reduced to an unobservable level by annealing
\cite{reppy2}.  It is by now very plausible that rapid freezing
produces highly disordered samples and that disorder is at the origin
of the experimental findings and of ``supersolid'' signals. Different
origins have been highlighted.  A recent work in
Ref.~\onlinecite{BPS2} suggested that the {\it core} of dislocations
could be superfluid.  A phenomenological theory of the role of
dislocations can be found in Ref.~\onlinecite{Toner}. Following up
Ref.~\onlinecite{PGG} a scenario to explain the experimental findings
in terms of dislocation induced superfluidity has been put forward in
Ref.~\onlinecite{bouchaudbiroli}. From the experimental point of
view, the role of grain boundaries has been analyzed in
Ref.~\onlinecite{balibar}. Another idea introduced in
Refs.~\onlinecite{BPS1,Nussinov} is that rapid freezing produces
amorphous glassy solids that become supersolid at low
temperature. This ``superglass'' phase has been detected in Path-Integral
Monte-Carlo (PIMC) numerical simulations\cite{BPS1}. 
These results are encouraging but addressing real-time dynamics
issues such as the stability and the dynamics of this phase
is out of reach of PIMC. 
Therefore, it is desirable to have a complementary analytical
approach that can give direct insight into these problems:
this is precisely the purpose of this work.

Our strategy will be to focus on a model that displays a superglass
phase and that can be concretely analyzed in some instances exactly
and in others using accurate and controlled approximations at zero
temperature. The model consists of bosonic particles interacting via a
short-range potential in three dimensions. Our approach is based on a
mapping between quantum Hamiltonians and (classical) Fokker-Planck
operators which allows us to obtain results on ground state properties
and time dependent correlation functions from the analysis of the
stochastic dynamics of a classical equilibrium system. This
connection, already well-known for a few decades \cite{ParisiBook,ZJ},
has been quite used and further studied in the context of the so
called Rokshar-Kivelson points
\cite{RoksharKivelson,Claudio,henley}. We will mainly focus on systems
of interacting bosons whose classical counterpart (via the above
mapping) is a system of hard spheres. These are known to undergo a
crystal\cite{Hoover} or a 
glass~\cite{KW87,MU93,Sp98,CFP98b,ReviewZamponi,PZ05,SK01,CKDKS} 
transition at high density. The glass transition takes place
when crystallization is avoided (either by fast compression or using different types 
of particles). 
The usefulness of the classical-quantum mapping is that it allows us
to obtain highly non-trivial, very accurate or, at instances, exact
results on the original quantum many-body problem by translating all
the knowledge on the Brownian dynamics of hard spheres. In particular
we will be able to work out in detail the properties of the
superglass, supercrystal and superfluid phases.
The drawback of this mapping is that the resulting quantum solids
are quite different from real He$^4$ crystals, mainly because of the
much smaller zero-point motion and of some peculiar properties of the
excitation spectrum to be discussed below.

The results of our work go beyond the physics of solid He$^4$ because
the superglass phase can emerge also in other physical contexts, in
particular strongly interacting bosonic mixtures of cold
atoms. Furthermore, another motivation for our study is to analyze a
quantum glassy phase not induced by quenched disorder. There are in
fact very few examples \cite{Chamon1,Chamon2,Tarzia,Ioffe,Schmalian} of that,
especially not relying on mean-field or large-$N$ approximations.

\section{Summary of the main results}
\label{Summary}
In the following section the mapping between classical Fokker-Planck
operators and quantum Hamiltonians is presented. In particular we will
obtain the relation between the classical potential and its quantum
counterpart. We will work it out for identical bosons as well as for
mixtures of bosons (mixtures may be relevant for cold atomic systems).
Furthermore, we shall introduce the model that we will focus on in
this paper, which is the quantum counterpart of classical Brownian
hard spheres. The quantum potential is still a hard sphere one but it
has in addition a sticky part at contact (and also a 3-body
interaction at contact).

In section IV we shall work out the zero temperature phase diagram of
the quantum model as a function of the density $\rho$. Actually we
will use instead of the density the packing fraction
$\phi=\pi\rho\s^3/6$ (where $\s$ is the sphere diameter) which is a
more natural variable for hard spheres. We find three possible phases:
superfluid, supercrystal and superglass. At low packing fraction the
system is superfluid but at $\phi\simeq 0.49$ it has a first order
transition toward a supercrystal phase. The transition toward the
superglass phase takes place at higher packing fractions, $\phi\simeq
0.62$. It can be achieved only if the transition toward the crystal is
avoided either by considering mixtures or compressing fast the liquid
phase. We will obtain quantitative results for the condensate
fraction. In particular for the superglass phase we will introduce and
compute an observable, akin to the Edwards-Anderson parameter for spin
glasses, that quantifies the inhomogeneity of the condensate
wavefunction in space. The values of the condensate fraction turn
out to be very small. This is likely due to the particular model we
have chosen which admits as ground state wavefunction the Gibbs
measure of hard spheres. However, exact results guarantee that,
although small, the condensate fraction is non-zero, {\it i.e.} it is
not an artifact. In Section V we focus on the real time quantum
dynamics close to the transition from the superfluid to the
superglass. In particular we obtain the behavior of the density
correlations as a function of time as well as the time dependent
condensate fluctuations.  Both correlation functions show a similar
behavior. Approaching the superglass from the superfluid, one finds
that after a rapid oscillating decay, there is a plateau indicating
frozen amorphous density and condensate fluctuations. On much larger
timescales these frozen fluctuations relax.  The transition to the
superglass phase takes place when the timescale for this second
relaxation goes to infinity (or, practically, becomes larger than any
relevant experimental timescale.). This behavior is actually very
reminiscent of the one of classical supercooled liquids approaching their
structural glass transition. In Section VI we will generalize our
model in order to make it more realistic. In fact, within the
classical-quantum mapping approach, the energy of the ground state is
always zero. In order to discuss the behavior of the pressure, the
order of the transitions and the superfluid properties we shall add a
small perturbation to the quantum potential originally introduced in
Section III. This will allow us to show that the transition from the
superfluid to the superglass is unusual to the extent that it is
thermodynamically first order but has also some characteristics of
second order transitions. Finally, conclusions, perspectives and
relations to experiments and numerical simulations will be presented
in Section VII.

\section{From Classical Langevin particles to zero temperature interacting 
Bosons}

The superglass is an amorphous quantum many-body state of interacting
bosons. To find such a state starting from a generic quantum
Hamiltonian of interacting particles is a daunting task. In this
paper, we take an approach that allows us to argue for such a phase in
a well controlled way, by constructing a {\it local} quantum many-body
Hamiltonian whose ground state is known exactly and can be argued to
be a glassy state at large enough densities. This formulation is
motivated by a generic result stating that for systems without a sign
problem there is a simple connection between quantum Hamiltonians and
the stochastic dynamics of a classical system. This can be used to
construct interesting Hamiltonians with real and non-negative ground
state wavefunctions related to classical equilibrium Boltzmann-Gibbs
measures. The connection is far-reaching since it allows one to obtain
controlled and highly non-trivial results on the phase diagram and
dynamical properties of a quantum many body problem~\cite{Claudio}.
Moreover, it has been used to construct an efficient algorithm
for Quantum Monte Carlo at zero temperature~\cite{Baroni}.

Here we shall explore this classical-quantum connection for the
particular case of bosonic point particles, following the standard
route for mapping classical Langevin dynamics for a many-particle
system and its associated Fokker-Planck operator to a
Schr{\"o}dinger operator~\cite{ParisiBook,ZJ}.

\subsection{Langevin dynamics: the Fokker-Planck and Schr{\"o}dinger operators}

Consider $N$ particles whose evolution is determined by the following
Langevin equations:
\begin{equation}
\label{Langevin}
\g_i \frac{d{\bf x}_i}{dt}=-\frac{\partial\;}{\partial \xx_i}\, U_N(\xx_1,\dots,\xx_N) 
+ \boldsymbol{\eta}_i(t) \ , \qquad i=1,...,N \ , 
\end{equation}
where $\g_i$ are friction coefficients, $\eta^{\alpha}_i(t)$ is a Gaussian white thermal noise with variance 
$\langle \eta^{\alpha}_i(t)\eta^{\beta}_j(t')\rangle=2T\,\g_i
\,\delta_{ij}\, \delta_{\alpha\beta}\,\delta(t-t') $. Furthermore, $T$ is the temperature (with $k_B=1$) and $\alpha$ and
$\beta$ run from $1$ to the spatial dimension $d$ (henceforth the boldface notation indicates vectors).  The potential
will eventually be assumed to be the sum of (symmetric) pair
potentials,
\begin{equation}\label{pot_gen}
U_N(\{\xx\})\equiv U_N(\xx_1,\dots,\xx_N) = \frac12 \sum_{i\neq j} V_{ij}(\xx_i-\xx_j) \ ,
\end{equation}
with $V_{ij}=V_{ji}$, and $\frac{\partial}{\partial \xx_i} U_N \equiv
\nabla_i U_N = \sum_{j (\neq i)} \nabla V_{ij}(\xx_i-\xx_j)$.

It is well known~\cite{ParisiBook,ZJ} that the evolution equation for the probability
distribution $P(\{\xx\})$ can be written as a Schr{\"o}dinger equation
in imaginary time:
\begin{equation}
\label{FP-imaginary-t}
\partial_tP=-H_{FP}P
\end{equation}
where the Fokker-Planck operator reads:
\begin{equation}\label{HFP}
H_{FP}=-\sum_i \frac{1}{\g_i} \frac{\partial}{\partial {\bf x}_i}\left[\nabla_i U_N +
T\frac{\partial}{\partial {\bf x}_i}  \right]
\end{equation}
The Fokker-Planck operator is non-Hermitian and can be proven to have
all eigenvalues larger than or equal to zero~\cite{ParisiBook,ZJ}. The zero eigenvalue
corresponds --as it can be readily checked-- to the stationary
distribution which is the equilibrium Gibbs probability measure:
\begin{equation}\label{GibbsP}
P_{\rm G}(\{\xx\})=\frac{1}{Z_N}\,e^{-\frac{1}{T}U_N(\{\xx\})}
=
\frac{1}{Z_N}\,e^{-\frac{1}{2T}\sum_{i \neq j}V_{ij}({\bf
x}_i-{\bf x}_j)}
\;.
\end{equation}

Setting for simplicity $\hbar=1$,
the Fokker-Planck operator can be mapped into a Hermitian quantum
Hamiltonian by a similarity transformation 
\begin{equation}
\label{H-HFP}
H=e^{\frac{1}{2T}U_N}\:
H_{FP} \:e^{-\frac{1}{2T}U_N}  ,
\end{equation}
that leads to
\begin{equation}\label{quantumH}
H=\sum_i \frac{1}{\g_i} \left[-T\frac{\partial ^2}{\partial {\bf x}_i^2}
-\frac{1}{2} \nabla^2_i U_N+
\frac{1}{4T} (\nabla_i U_N)^2
\right] = \sum_i \frac{{\bf p}_i^2}{2 m_i} + \VV_N(\{\xx\})
\end{equation}
This expression corresponds to a Hamiltonian for particles with 
mass $m_i = \g_i /(2T)$ and an effective potential which is the sum
of a two body and three-body interaction:
\begin{equation}\label{effectivepotential}
\begin{split}
\VV_N(\{\xx\}) &=
\sum_i \frac{1}{\g_i} \left[-\frac{1}{2} \nabla^2_i U_N+
\frac{1}{4T} (\nabla_i U_N)^2\right] \\ &= 
-\frac{1}{2}\sum_{j \neq i}\frac{1}{\g_i} \nabla ^2  V_{ij}({\bf x}_i-{\bf
  x}_j)+\frac{1}{4T} \sum_{i; j (\neq i); j' (\neq i)} \frac{1}{\g_i}
\nabla V_{ij}({\bf x}_i-{\bf x}_j) \cdot \nabla V_{ij'}({\bf x}_i-{\bf
  x}_{j'})
\;.
\end{split}
\end{equation}
The eigenfunctions of the quantum Hamiltonian and of the Fokker Planck
operator are in a one to one correspondence.  Indeed, by applying the
similarity transformation introduced above one finds:
\begin{equation}\label{WF}
\Psi_E(\{\xx\}) \propto e^{\frac{U_N}{2T}} \;P_E(\{\xx\})
\end{equation}
where $P_E$ indicates the right eigenfunction of the Fokker-Planck
operator with eigenvalue $E$, and $\Psi_E$ its counterpart associated
to the quantum Hamiltonian.  This also implies that all the
eigenvalues $E$ corresponding to the Fokker-Planck operator are
identical to the ones of the quantum Hamiltonian.

In particular, this relation, together with Eq.~(\ref{GibbsP}),
allows one to obtain straightforwardly the ground
state wavefunction of the quantum problem, which is of the Jastrow
form~\cite{Jastrow,McMillan,FCR70}:
\begin{equation}\label{Jastrow}
\Psi_G(\{\xx\}) =\sqrt{P_{\rm G}(\{\xx\})}
= \frac1{\sqrt{Z_N}} 
\exp \left[-\frac{1}{4T}\sum_{i \neq j}V_{ij}({\bf x}_i-{\bf  x}_j) \right] \ .
\end{equation}

The logic of the approach we pursue hereafter is the following: we
take as starting point Hamiltonians with many-body potentials of the
form Eq.~(\ref{effectivepotential}), for which the Jastrow form
Eq.~(\ref{Jastrow}) is exact. In general, wavefunctions of this form
lead to more than two-body interactions $\VV_N(\{\xx\})$ (note that
also He$^4$ has weak higher order interactions). The important point
is that if the two-body potentials $V_{ij}({\bf x}_i-{\bf x}_j)$ are
short-ranged ({\it i.e.} local) potentials, then $\VV_N(\{\xx\})$ is
also local, and thus the many-body Hamiltonians on which we focus are
{\it local} (non-local Hamiltonian may lead to pathological
behaviors).

Because we know exactly the ground state wavefunction, and it
is related to a Boltzmann-Gibbs measure for a classical system,
quantum static correlation functions can be computed in terms of
classical static correlation functions~\cite{FCR70}. Furthermore, as we shall show 
and noticed by Henley~\cite{henley}, the mapping
generalizes also to dynamical correlation functions.  Hence, 
we will obtain quantum dynamical correlation functions at zero temperature 
by analytic continuation of classical (stochastic) dynamical
correlation functions. 

To simplify the notations, in the following we will fix
$T=1$ and $\la \g \ra = N^{-1} \sum_i \g_i =1$. Together with
$\hbar=1$, this fixes the units in both classical and quantum
problems.
As a consequence the masses in the quantum problem read 
$m_i=\frac12 \frac{\g_i}{\la \g \ra}$.

\subsection{Identical bosons}

Let us first consider the simplest case of $N$ identical {\it bosons}
characterized by the Hamiltonian $H$ in Eq.~(\ref{quantumH}) with
$\g_i \equiv \g =1$ and $V_{ij} \equiv V$.  It is important to remark
that since the particles are bosons one has to consider only many-body
states that are completely symmetric under permutation of particles,
and study {\it only} observables that are invariant under permutation
of particles (e.g. the density-density correlator).  This is clearly
not a difficult constraint to handle since the Jastrow form
(\ref{Jastrow}) with $V_{ij}=V$ is completely symmetric. Furthermore,
even in the study of dynamical correlations (section~\ref{sec:5}) this
will not be a problem because if one starts from a probability law
completely symmetric under permutation of particles, the
symmetrization carries over to all later times. This follows trivially
from the Fokker-Planck evolution Eq.~(\ref{FP-imaginary-t}), since if
the state $P$ and the operator $H_{FP}$ are both symmetric under
exchange of particles, so is the time derivative $\partial_t P$ and
thus the many-body state thereafter.

For a given classical isotropic 2-body potential $V(\xx)=V(|\xx|)$,
the resulting quantum potential energy will have 2-body and 3-body
interactions:
\begin{eqnarray}
\nonumber
\VV_N(\{\xx\})&=&\sum_{i>j} v^{\rm pair}(\xx_i-\xx_j)+
\sum_{i \neq j \neq j' \neq i} v^{\rm 3-body}(\xx_i-\xx_j,\xx_i-\xx_{j'})\\
v^{\rm pair}(\xx)&=&-\nabla ^2  V(\xx)+\frac{1}{2}\,  
[\nabla V(\xx)]^2
= -\frac{d-1}{r}\,V'(r)
-V''(r)
+\frac{1}{2}\, [V'(r)]^2
\label{eq:pair}
\\
v^{\rm 3-body}(\xx,\xx')&=&\frac{1}{4}  \;
\nabla V(\xx) \cdot \nabla V(\xx')
\nonumber
=\frac{1}{4}\,
\frac{\xx}{r}\cdot\frac{\xx'}{r'}
\;V'(r)\,V'(r')
\;,
\end{eqnarray}    
where $d$ is the spatial dimension and $r \equiv|\,\xx|$.

\subsection{Binary mixtures of bosons}

It is also worthwhile considering binary mixture of bosons. In fact
the superglass phase we shall discuss in the following will emerge
only when crystallization is avoided.  We expect, as it is the case
for classical systems, that this can be obtained either by fast
compression of the liquid state or by considering mixtures of bosons.
Mixtures typically induce a frustration effect on the crystalline
phase whose energy, as a consequence, increases. Instead the glass
phase, thanks to its disordered structure, is less affected and,
hence, starts to compete in terms of energy (or dynamical basin of
attraction) with the crystalline phase. This conclusion, which we
expect to be correct on physical grounds, is also suggested by
approaches based on variational wavefunctions of the Jastrow form
Eq.~(\ref{Jastrow}). These types of wavefunctions map the quantum
problem into a classical one and are very effective in describing a
number of the observed phenomena in the quantum Bose
fluid. Lennard-Jones type potentials (as classical potentials in the
exponent of the Jastrow wavefunction) have been considered by
McMillan~\cite{McMillan}, Francis, Chester and Reatto~\cite{FCR70} in
the study of He$^4$ superfluids. For a single species of particles,
such interaction potentials lead to classical liquid states and, at
very high densities, might lead to a crystalline state~\cite{HL68}.
Within Jastrow-like approaches, thanks to this analogy with classical
systems, we expect the physical effect induced by mixtures to be
similar to what happens in classical liquids.

As a consequence, we conjecture that for a binary mixture, a quantum
glassy state may emerge at high enough densities. Experimentally, the
binary mixture can be obtained using two species of bosonic atoms, $A$
and $B$. It may be the case that we do not need to develop a
condensate fraction for both species $A$ and $B$; it would suffice to
have a superglass of, say, species $A$, while species $B$ always
remains normal, with its role solely that of enabling the formation of
the glassy state. These type of binary mixture can be obtained in cold
atom experiments.

The generalization of the results of the previous section to a binary
mixture is presented in Appendix~\ref{sec:appendix}. In the following
we will focus on the identical boson system but one must keep in
mind that to stabilize the glassy phase one may have to consider
mixtures or fast compression of the liquid state. Thus, our strategy will be to
study directly the glass phase of identical particles and translate
the results to the mixture case where this phase actually exists.

\subsection{Quantum model with hard sphere wavefunction}

Let us investigate what quantum bosonic system in particular
corresponds to a classical hard sphere problem upon the mapping
above. The motivation for this study is that the classical hard sphere
packing is a well studied problem in the context of glasses for large
enough packing fractions~\cite{KW87,MU93,Sp98,CFP98b,ReviewZamponi,PZ05,SK01,CKDKS}, 
and we will be able to use these results for
substantiating the notion of the superglass state. We shall focus on a
classical potential $V(r)=V_0 \exp(-\lambda [(r/\s)^2-1])$, where
$\sigma$ is the characteristic sphere size, perform the mapping and
take the $\lambda\to\infty$ limit that enforces the hard sphere
constraint. Note that although in principle the form of the potential
is not very important as long as it is infinite for $r<\s$ and zero
for $r>\s$ (in the $\lambda\to\infty$ limit), the previous form makes
the following discussion particularly easy.  We also set $V_0=1$ and
$\sigma=1$ for simplicity:
indeed, even if the units have already been fixed, 
$V_0$ is irrelevant in the $\l\to\io$
limit, while a change in $\sigma$ is 
equivalent to a change in the density.

\begin{figure}
\includegraphics[width=8.5cm]{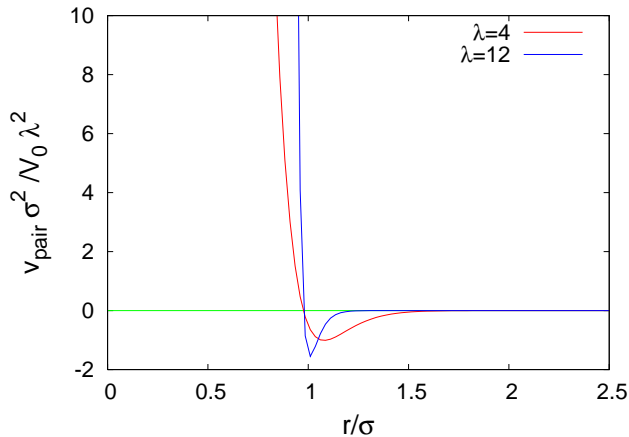}
\caption{Form of the pair potential $v^{\rm pair}(r)$ in the mapped
quantum problem that derives from the classical potential $V(r)=V_0
\exp(-\lambda [(r/\s)^2-1])$. (Notice that we have set $V_0=1$ and
$\s=1$ in the text.)}
\label{fig:pair-potential}
\end{figure}

Let us first discuss the pair term, which depends only on the
interparticle distance $r$:
\begin{equation}\label{pairpotential}
v^{\rm pair}(r)=[2\lambda d-4
\lambda^2 r^2 ]\;V(r)+2\lambda^2 r^2\;[V(r)]^2
\;,
\end{equation}
where we have used explicitly the (convenient) exponential form of the
potential. In Fig.~\ref{fig:pair-potential} we show the form of this
two-body potential.

The form of the pair potential in the large $\lambda$ limit is simple:
\begin{itemize}
\item For $r>1$, the pair potential is zero because the exponential terms
go very fast to zero. 
\item For $r<1$, the pair potential goes to infinity. Although the second term 
is negative, the third one is much larger than the first two and dominates. 
\item Very close to $r=1$, within a window that shrinks to zero
in the large $\lambda$ limit, the potential becomes negative and goes
to minus infinity. Indeed, for $r=1$, its value is
$2d\lambda-2\lambda^2$.  The potential remains attractive (negative)
on an interval of size $1/\lambda$. This attractive part of the
potential is responsible for the fact that ground state wavefunction
does not vanish (or is discontinuous) at the contact between hard
spheres. Particles can stay close to contact paying a lot of kinetic
energy, ${\OO}(\lambda^2)$, and gaining a lot of potential energy,
also $\OO(\lambda^2)$. \\

\end{itemize} 

Let us now consider the 3-body term. It can be written as:
\begin{equation}
v^{\rm 3-body}(\xx,\xx')=\lambda^2\; \xx \cdot \xx'  \;
V(r) \,V(r') \ ,
\end{equation}
where for a given triplet of particles $i,j,j'$ one has
 $\xx = \xx_i - \xx_j$ and $\xx'= \xx_i - \xx_{j'}$, see Eq.~(\ref{eq:pair}).
The form of the 3-body potential in the large $\lambda$ limit is
simple: it is non-zero only if {\it both} particles $j$ and $j'$ have a non-zero overlap with particle $i$. When there is a finite overlap (in the large $\lambda$ limit) 
the contribution coming from the 3-body term, which can be actually positive or negative, is always smaller
than the one coming from the sum of the pair contributions
$ij$ and $ij'$. Thus, it can be neglected. The outcome coming from the pair interactions is that
all particle configurations 
with finite particle overlaps are simply 
excluded by the Hilbert space.
When both $r$ and
$r'$ are very close to $1$, of the order $1/\lambda$, the 3-body
term will give rise to a non-negligible contribution similar to the one studied for
the pair contribution. The main difference is that it can be
attractive or repulsive depending on the relative orientation of $\xx$
and $\xx'$. As a conclusion the interaction between particles is the hard sphere one plus a contact term that 
is sticky for the pair contribution and that can be repulsive or attractive for the triplet term depending 
on the geometry of the triplet.

\section{Superfluid, supercrystal and superglass phases}

Here we make use of the relationship between the classical and
quantum models, and study the zero temperature phases of the quantum
system introduced in the previous section III.D. We will work out
 the properties of the quantum phases by using the
classical Gibbs measure defined by the square of the Jastrow
wavefunction~\cite{Claudio,McMillan,FCR70,HL68}.

\begin{figure}
\includegraphics[width=8.5cm]{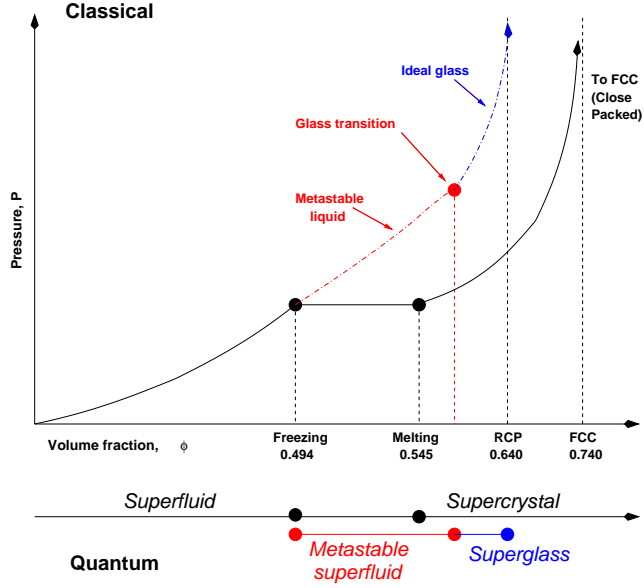}
\caption{Classical phase diagram for hard sphere systems (top) and its quantum counterpart
(bottom).
}
\label{fig:phasediagram}
\end{figure}

We shall focus on a system of $N$ identical bosons
with the particular interaction, discussed in the previous section,
that corresponds to a Jastrow state with an hard sphere potential. Having
fixed $\s=1$, the
control parameter in this case is the particle density $\r = N/V$ of the
particles, or the {\it packing fraction} $\phi=\pi \r/6$.
Let us recall first the phase diagram of classical hard spheres\cite{Hoover,PZ05,phasediagramhardspheres} (see top 
of Fig.~\ref{fig:phasediagram}). 
At low density, the system is liquid, and upon increasing
the density it undergoes a first order phase transition to a
crystalline state, that is arranged in a face centered cubic
(FCC) lattice. Moreover, a metastable glassy phase can be obtained in
the classical problem if the density is increased fast enough or in presence
of small bi-dispersity (binary mixtures). 
This glass phase can be compressed until the random close packing
packing (RCP) fraction $\phi\sim 0.64$. These three
classical phases have their corresponding counterparts in the
mapped quantum system. As we shall show, these are superfluid, supercrystal
and superglass phases.  
The classical phase diagram and its quantum counterpart
are shown in Fig.~\ref{fig:phasediagram}.

Below we discuss the property of these three quantum phases in some detail
with particular emphasis on their one-particle density matrix and Bose
condensate fraction. We will treat the glass phase as a true
equilibrium phase because, as we have already discussed,
it will emerge when crystallization is avoided either by fast compression 
of the liquid state or considering suitable binary mixtures.

\subsection{Off-diagonal long range order and classical correlation functions}

Let us first recall general results on off-diagonal long range
order (ODLRO)~\cite{Onsager,Leggettbook}. In particular, 
we shall show that for Jastrow-type wavefunctions there is 
a simple relation~\cite{Onsager,McMillan,FCR70} 
between the structure factor of a classical
system and the one particle density matrix. This relation has a particularly 
simple form in the case of hard spheres. 
The condensate fraction is related to the amount of
off-diagonal long range order in the quantum problem and can be therefore
related to an observable for the classical corresponding liquid. In
appendix~\ref{appendix:binary-ODLRO} we generalize the following  discussion of
ODLRO to the case of binary mixtures.

The one particle off-diagonal density matrix reads in term of the ground state wavefunction $\Psi_G$ ($V$ is the
volume): 
\begin{equation}\label{Rdef}
\RR({\xx},{\xx'})=V\int d{\xx}_2...d{\xx}_N\:
\Psi_G({\xx},{\xx}_2,...,{\xx}_N)\:
\Psi_G({\xx'},{\xx}_2,...,{\xx}_N) = V \sum_i n_i \;\psi_i({\xx})\,\psi^*_i({\xx}') 
\end{equation} 
where the $n_i$ and $\psi_i({\xx})$ are, respectively, eigenvalues and
(orthonormal) eigenvectors of the one particle density matrix
$\RR(\xx,\xx')$, $\int d{\xx'}\, \RR({\xx},{\xx'})\,\psi_i({\xx}')=
n_i\,\psi_i({\xx})$, see e.g. Ref \onlinecite{Leggettbook}.  One can
interpret $n_i$ as the fraction of particles in state $i$, and $\sum_i
n_i = 1$.  The largest eigenvalue $n_0$ gives directly the condensate
fraction.

If $\Psi_G$ is given by a Jastrow wavefunction with a hard sphere
potential, this expression is particularly simple in terms of a
classical equilibrium correlation functions:
\begin{equation}
\RR({\xx},{\xx'})=\frac{V}{N(N+1)} \frac{Z_{N+1}}{Z_N}
\;e^{V({\xx}-{\xx}')/T} 
\;\sum_{i\neq j} \langle\delta({\xx}-{\xx}_i) \, \delta({\xx}'-{\xx}_j) \rangle =
\frac{Z_{N+1}}{Z_N (N+1)} \,\rho^{-1}\, \rho({\xx}) \,\rho({\xx}')\; 
y({\xx},{\xx}')
\, ,
\label{eq:ODLRO-hard-spheres}
\end{equation}
where $\rho({\xx}) = \langle \sum_i \delta({\xx}-{\xx}_i) \rangle$ and
$y({\xx},{\xx}') = e^{V({\xx}-{\xx}')/T} \,g({\xx},{\xx}')$, and
$g({\xx},{\xx}')$ is the usual pair correlation 
function~\cite{HansenMacDonald}. Notice
that this expression follows by considering a system of $N+1$
particles, with two of them fixed at $\xx$ and $\xx'$, and that the
difference between $y({\xx},{\xx}')$ and $g({\xx},{\xx}')$ arises
because $\RR({\xx},{\xx'})$ does not vanish when $\xx$ and $\xx'$ are
less than a particle diameter away. For hard spheres, there is a
simplification and $g({\xx},{\xx}')=y({\xx},{\xx}')$ for $|\xx-\xx'|>1$;
however, for $|\xx-\xx'|<1$ the pair correlation function vanishes 
while $y(\xx,\xx')$ is finite~\cite{HansenMacDonald}.

The prefactor in Eq.~(\ref{eq:ODLRO-hard-spheres}) is related to the
fugacity of the system and it is easily evaluated in the large $N$
limit:
\begin{equation}
Z_N \sim N! \, e^{ N S(N/V) } \ \ \ \Rightarrow \ \ \ 
\frac{Z_{N+1}}{Z_N (N+1)} = e^{\frac{d}{d\rho}[ \rho\, S(\rho)]} = 
\rho^{-1} e^{\frac{d}{d\rho}[ \rho \,S_{\rm ex}(\rho)]} 
\equiv \rho^{-1} f(\rho) = \frac1{z(\rho)}
\label{eq:f-def}
\end{equation}
where $S(\rho)=1-\log \rho + S_{\rm ex}(\rho)$ is the entropy of
classical hard spheres as a function of the average density $\rho =
N/V$, $S_{\rm ex}(\rho)$ is the excess entropy with respect to the
ideal gas, and $z(\rho)$ is the fugacity.  Using this relation we find
the final, quite simple, expression:
\begin{equation}\label{Rnoninv}
\RR({\xx},{\xx}')= f(\rho) \,\frac{\rho({\xx})\,\rho({\xx}') }{\rho^2}\;
y({\xx},{\xx}')
\,.
\end{equation}
For a homogeneous phase, $\r({\xx})=\rho$ and $\RR({\xx},{\xx}')$
becomes a function of $|\xx-\xx'|$ only,
\begin{equation}\label{Rinv}
\RR(|{\xx}-{\xx}'|)= f(\rho) \,y(|{\xx}-{\xx}'|)
\end{equation}
and the condensate fraction is just $\lim_{|{\xx}-{\xx}'|\to\infty}
\RR(|{\xx}-{\xx}'|) = f(\rho)$, since $y({\xx},{\xx'})\to 1$ (like the
pair correlation function) as $|{\xx}-{\xx'}|\to\infty$. This
expression shows that as long as the fugacity is finite, {\it i.e.}
the pressure is finite, we shall find a non-zero condensate
fraction\cite{Onsager}. As a consequence, we expect a superfluid phase at low
densities, then a first order transition into a supersolid crystalline
phase as the density is increased. In the case of non-identical hard
spheres, e.g. bi-disperse, or for extremely rapid quenches, the
system should end up in a superglass phase at high density, and the
condensate fraction will vanish only at close packing.
In the following, we will apply the results above and 
study in detail the superfluid, supercrystal and superglass phase.

\subsection{Superfluid phase}

\begin{figure}
\includegraphics[width=8.5cm]{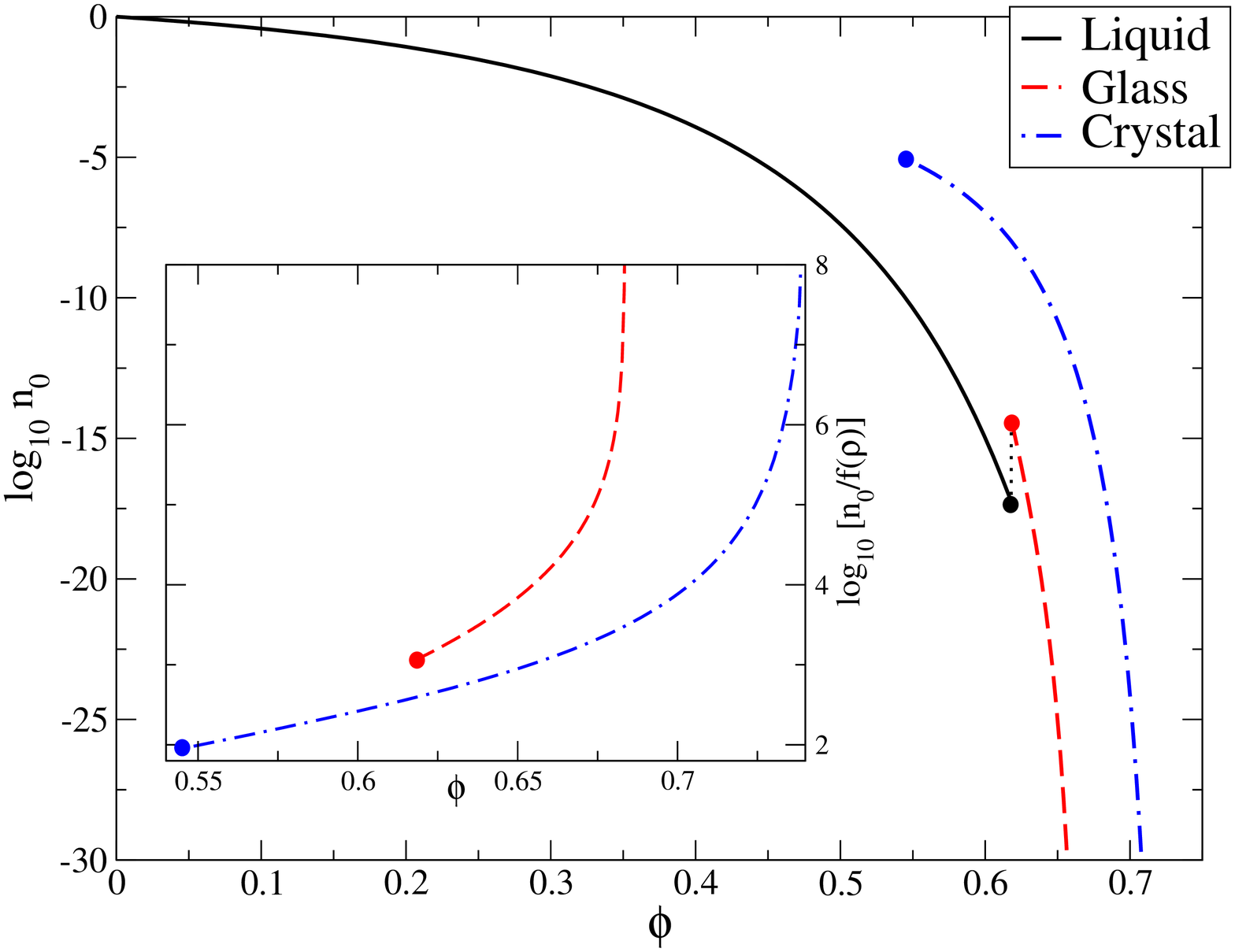}
\caption{Condensate fraction as a function of the packing fraction
for the hard sphere Jastrow wavefunction.
Full (black) curve: superfluid phase, Eq.~(\ref{n0liq}).
Dot-dashed (blue) curve: supercrystal phase, Eq.~(\ref{n0cry}).
Dashed (red) curve: superglass phase, Eq.~(\ref{n0gla}); note that
the theory slightly overestimates the random close packing density
with respect to the commonly accepted value of $\phi=0.64$.
The blue dot marks the location of the melting transition, where 
liquid-crystal phase coexistence begins, while the red dot marks the glass
transition density. Inset: the enhancement factor $n_0/f(\r)$ due to
the spatial inhomogeneity of the crystal and glass phases. Note that
this factor is quite large, of the order of $10^2-10^3$ at the melting
and glass transition densities. Still, the condensate fractions we find
are extremely small, probably due to the
classical-like nature of solids with the Jastrow hard sphere
wavefunction.
}
\label{fig:m}
\end{figure}

At low packing fraction the classical system is in the liquid phase. 
Analogously, the quantum system is in a quantum liquid phase. 
As discussed above, we expect it to be superfluid. In the following we 
shall compute its corresponding condensate fraction. In order
to obtain a heuristic semi-quantitative expression for the condensate
fraction, we used the Carnahan-Starling
approximation~\cite{HansenMacDonald} for the classical hard sphere system.
This is a kind of virial
resummation known to work very well for describing the properties of
the hard sphere system in the whole fluid range. Within this
approximation one finds \cite{HansenMacDonald}
\begin{equation}
\rho\, S_{\rm ex}(\rho) = 
\frac 1 V \log \left(Z_N/V^N\right)=
-\frac{6\phi}{\pi}  \frac{\phi (4-3\phi)}{(1-\phi)^2}  
\;.
\end{equation}
Then Eq.~(\ref{eq:f-def}) leads to the following expression for the condensate fraction $n_0$:
\begin{equation}
\label{n0liq}
n_0 = f(\r) =\exp \left[\frac{\phi(-8+9\phi-3\phi^2)}{(1-\phi)^3} \right]
\;.
\end{equation}
This is plotted in Fig.~\ref{fig:m}. Note that
the freezing crystallization transition takes place at a
packing fraction $\phi\simeq 0.494$ for hard spheres, 
see Fig.~\ref{fig:phasediagram}.

\subsection{Crystal phase}
For $\phi> 0.545$, the classical hard sphere system is in the crystal phase
(for $0.494 < \phi < 0.545$ there is phase coexistence between the liquid and
the crystal). 
Correspondingly, the quantum system displays a quantum crystal phase. 
In the following we shall compute its condensate fraction. 
The crystal phase is inhomogeneous but still 
$y({\xx},{\xx'})\to 1$ at large $|{\xx}-{\xx'}|$;
therefore we have\cite{Onsager}
\begin{equation}
\lim_{|{\xx}-{\xx}'|\to\infty} \RR({\xx},{\xx}') 
= f(\rho)\; \frac{\rho({\xx})\,\rho({\xx}') }{\rho^2} = n_0 \, V
\;  \psi_0({\xx}) \, \psi^*_0({\xx'})
\;,
\end{equation}
and from this factorization
one obtains that
$\sqrt{n_0\,V}\;\psi_0(\xx)=\sqrt{f(\rho)}\;\rho({\xx})/\rho$ (notice
that the eigenvector $\psi_0(\xx)$ is real valued, as $\RR({\xx},{\xx}')$
is real and symmetric). It thus follows from the normalization of
$\psi_0(\xx)$ that
\begin{equation}
 n_0 = f(\rho) \;\frac{1}{\rho^2} \frac{1}{V}\int d{\xx} \;{\rho({\xx})}^2 \ .
\end{equation}
A rough estimate of $n_0$ is given by $f(\rho)$ since 
$f(\rho)
\leq n_0  
\leq
\left( \max_{V} \frac{ \rho({\xx})}{\rho} \right)^2 f(\rho)$. 
The first inequality follows from $\rho^2=\left[\frac{1}{V}\int d{\xx}
\;{\rho({\xx})}\right]^2\leq \frac{1}{V}\int d{\xx} \;{\rho({\xx})}^2$.
However, because of the inhomogeneity of the density profile, 
the contribution to $n_0$ coming from the integral of the density squared
may be quite large. 
In order to estimate it we assume that
\begin{equation}
\rho({\xx}) = \sum_i \frac{e^{-\frac{|{\xx}-{\bf R}_i|^2}{2 A}}}{(2\pi
  A)^{3/2}}
\end{equation}
where ${\bf R}_i$ are the FCC lattice sites and the Gaussians are
roughly non-overlapping. It has been shown by numerical simulations that
this approximation is rather good\cite{YA74} and values of $A$ have been
computed by density functional theory\cite{DAC95}.
Using these results one finds
\begin{equation}
\label{n0cry}
n_0 = f(\r) \frac{1}{\rho^2} \frac{1}{V}\int d{\xx} \;{\rho({\xx})}^2 =
 \frac{f(\r)}\rho \frac1{(4\pi A)^{3/2}}
\;.
\end{equation}
with $f(\r)$ given in (\ref{eq:f-def}).

To compute $f(\r)$ we used the phenomenological equation of state for the 
FCC crystal phase of classical hard spheres proposed by Speedy~\cite{Speedy}:
\begin{equation}
\frac{P}{\r} = \frac{3}{1-z} - a\frac{z-b}{z-c} \ ,
\hskip2cm z = \phi/\phi_{\rm FCC} \ ,
\end{equation}
where $\phi_{\rm FCC} = \pi/(3 \sqrt{2}) = 0.74\ldots$, $a=0.5921$,
$b=0.7072$, $c=0.601$. The entropy can be obtained by integrating this
relation with respect to density as detailed in
Ref.~\onlinecite{Speedy} where the integration constant is also
reported.  Values of $A$ have been taken from table I of
Ref.~\onlinecite{DAC95}. They have been fitted to a polynomial to get
\begin{equation}
\sqrt{A} = a_1 \, (\phi_{\rm FCC}-\phi)+
a_2 \, (\phi_{\rm FCC}-\phi)^2+
a_3 \, (\phi_{\rm FCC}-\phi)^3 \ ,
\end{equation}
with $a_1=0.35$, $a_2=-0.866$ and $a_3=3.58$.  The final result for $n_0$ is
reported in Fig.~\ref{fig:m}.

It is worth to note that the crystal phase is not commensurate, except
at close packing. Therefore our results are consistent with the general
statement by Prokofev and Svistunov, that commensurate crystals are not
supersolids\cite{PSTh}.

\subsection{Glass phase}
\label{sec:glass-phase}

In order to understand the superglass phase, one must first translate
the results of the classical hard sphere problem into the quantum
one. Thus, let us start by discussing the classical 
results. 

It is well known that the classical hard sphere problem may undergo
a glass transition at a packing fraction $\phi_K \sim 0.6$ if
compressed sufficiently fast~\cite{KW87,CFP98b,ReviewZamponi,PZ05,phasediagramhardspheres} or
in case of mixture of different particles~\cite{SK01,CKDKS}. This has
been found analytically under some
approximations~\cite{KW87,MU93,CFP98b,ReviewZamponi,PZ05}, in
simulations~\cite{SK01,SDST06} and in experiments on colloidal
systems~\cite{PM87,WCLSW}.  The classical phase diagram is reported in
Fig.~\ref{fig:phasediagram}.

The physical mechanism behind this transition in finite dimension is
still unclear.  However, approximated theories able to provide
good quantitative predictions have been developed. The most successful
is the so-called Mode-Coupling Theory (MCT)~\cite{MU93,phasediagramhardspheres,Go99} that
correctly describes the dynamics of the Langevin system
Eq.~(\ref{Langevin}) for densities slightly smaller than the glass
transition density. A complementary approach is based on the replica
method~\cite{MP99b,CFP98b,ReviewZamponi,PZ05} and gives predictions on
static observables at and above the glass transition density. We will
discuss in the following what predictions can be derived from these
theories for the quantum problem.

Our strategy will consist in deriving results for the quantum glassy
phase starting from the known results for the classical glassy phase.
We will also provide quantitative results using the replica approach
which has been shown to give a reasonable description of numerical
simulation results~\cite{ReviewZamponi,PZ05}.  Our working hypothesis
is that an ideal classical glass transition indeed takes place: at low
density (or packing fraction) the system is in the liquid phase and
above a critical density (assuming that crystallization is avoided)
there is a thermodynamic transition toward an amorphous state of
matter which is the glass state. In the glassy state the system can
be frozen in many different amorphous configuration or states. Note
that assuming that the classical glass transition is a real
thermodynamic phase transition just simplifies the presentation. In
fact, even if this is not the case in reality, one can translate and
generalize all the following discussion: amorphous thermodynamic
states will then become just metastable amorphous configurations in
which the system is trapped on the relevant experimental timescales.

\subsubsection{The decomposition of the Gibbs measure in pure states}

Following the strategy outlined above, we will therefore assume that
at the (classical) glass transition, the Gibbs measure $P_G(\{\xx\})
=|\Psi_G(\{\xx\})|^2$ splits into a very large number of thermodynamic
states $P_\alpha(\{\xx\})$:
\begin{equation}\label{decomposition}
|\Psi_G(\{\xx\})|^2 = \sum_\alpha w_\alpha\; P_\alpha(\{\xx\}) \ .
\end{equation}
An operative description of $P_\alpha(\{\xx\})$ is the Boltzmann
measure obtained by coupling the system to an infinitesimal
non-homogeneous external potential, $V_\alpha$, that forces the system
into the state $\alpha$:
\begin{equation}\label{pinning}
P_\alpha(\{\xx\}) = \lim_{\epsilon \rightarrow 0}
\frac{e^{-U_N/T-\epsilon V_{\alpha}/T}}{Z_\alpha} \ ,
\hskip1cm
Z_\a =  \lim_{\epsilon \rightarrow 0} 
\int d\{\xx\} e^{-U_N/T-\epsilon V_{\alpha}/T} \ .
\end{equation}
Note that clearly the potential $V_\a$ must be chosen symmetric under
particle exchanges. From the above definition it follows 
that the states $P_\a$ are also completely Bose symmetric like the full Boltzmann
measure $P_G$.

The tricky aspect of glassy systems is that this external potential
(or field) is not known a priori as it is for, say, the ferromagnetic
case.  The reason is that it is random as the state it
pins. $w_\alpha$ is the thermodynamic weight of the state $\alpha$.
It equals $Z_\alpha/Z_N$, where $Z_\alpha$ is the partition function
in the presence of an infinitesimal pinning field forcing the system into
the state $\alpha$. $Z_N$ is instead, as before, the partition
function in the absence of any pinning field.  Hence, it contains the
contribution of all different states: $Z_N=\sum_\alpha Z_\alpha$ and
then $\sum_\a w_\a = 1$.

Since the $P_\alpha(\{\xx\})$s are equilibrium steady state
distributions one can obtain an eigenfunction with minimum (zero)
energy using the similarity transformation introduced in
Eq.~(\ref{WF}):
\begin{equation}\label{psistate}
\Psi_\alpha(\{\xx\})={\cal N}_\alpha\; \sqrt{Z_N}\;e^{U_N/2T}\;P_\alpha(\{\xx\}) =
\NN_\alpha\; P_\alpha(\{\xx\})/\Psi_G(\{\xx\}) \ .
\end{equation}
${\cal N}_\alpha$ is a normalization constant that has to be fixed by
imposing that $\Psi_\alpha(\{\xx\})$ is normalized:
\[
1={\cal N}_\alpha^2\;\lim_{\epsilon \rightarrow 0}\int d\{\xx\} \;
Z_N\; e^{U_N/T}\;\frac{e^{-2U_N/T-2\epsilon V_{\alpha}/T}}{Z_\alpha^2}= 
 {\cal N}_\alpha^2\;\lim_{\epsilon \rightarrow 0}\frac{Z_N}{Z_\alpha^2}\int d\{\xx\} 
\;e^{-U_N/T-2\epsilon V_{\alpha}/T}=
{\cal N}_\alpha^2\frac{Z_N}{Z_\alpha}
\] 
Note that we have used explicitly that one always recovers $Z_\alpha$
whether one uses $\epsilon$ or $2\epsilon$. The outcome is that 
${\cal N}_\alpha=\sqrt{Z_\alpha/Z_N}=\sqrt{w_\alpha}$.  Using this
result and plugging the expression (\ref{psistate}) into
(\ref{decomposition}) one finds:
\[
\Psi_G(\{\xx\})^2 = \Psi_G(\{\xx\}) \sum_\alpha \sqrt{w_\alpha}
\;\Psi_\alpha(\{\xx\})  \ .
\]
By factoring out $\Psi_G(\{\xx\})$ one finally finds~\footnote{
It is worth
to stress that in general Eq.~(\ref{psieq}) is {\it not} a decomposition in pure states
of a Gibbs measure like Eq.~(\ref{decomposition}). 
The crucial difference is that in Eq.~(\ref{decomposition}) the $P_\a$ are
normalized to 1 and the weights $w_\a$ add up to one, while this is {\it not} the
case in Eq.~(\ref{psieq}).
}:
\begin{equation}\label{psieq}
\Psi_G(\{\xx\}) = \sum_\alpha \sqrt{w_\alpha}\;
\Psi_\alpha(\{\xx\})  \ .
\end{equation}
The ground state wavefunction is therefore the coherent sum of the
wavefunctions corresponding to states $\alpha$. 

However, no interference is present in the thermodynamic limit and all
cross-terms in the square of the wavefunction can be dropped. This can be
shown as follows: first, by plugging the expression of ${\cal N}_\alpha$ into
Eq.~(\ref{psistate}) one finds that
$\Psi_\alpha(\{\xx\})=\sqrt{P_\alpha(\{\xx\})}$.  Using this result
and noticing that the square of the previous equation (\ref{psieq})
has to give back Eq.~(\ref{decomposition}) we find that the
interference (cross-products) terms have to be zero.  This result can
be understood in a simple heuristic way: configurations on which the
wavefunction $\Psi_\alpha$ is concentrated have extremely small weight
in any other $\Psi_\beta$.  Any given interference term $\alpha,\beta$
is expected to give a contribution decreasing exponentially fast in
$N$. Since the number of states, hence the number of couples
$\alpha,\beta$, increases at most exponentially in $N^{(d-1)/d}$ (see
Ref.~\onlinecite{NewmanStein}) one finds that the interference terms
can be neglected in the thermodynamic limit.

Thus, the probability of finding the quantum system in a
state $\a$ is not different from that of the classical
problem: both are given by $w_\a$ in absence of a pinning external potential.
The pinning potential $V_\a$ introduced in Eq.~(\ref{pinning}) induces a corresponding
pinning potential ${\cal V}_\a$ via the mapping introduced in 
Eq.~(\ref{effectivepotential}). The effect of the pinning potential
is to concentrate the Jastrow wavefunction on state $\a$; equivalently,
in the quantum problem, the pinning potential ${\cal V}_\a$ lifts the
degeneracy between quantum ground states and selects $\Psi_\a$ as the
unique ground state of the system.
As a consequence each one of the $\Psi_\alpha$s has to be interpreted as a possible
state of the system (a more direct dynamical interpretation that shows that the system 
does not
escape by tunnelling once it is in a state $\alpha$ is presented in Sec.~V) .
The glass phase in the quantum system is thus a random solid
in the same sense as its classical counterpart.
However, much as in
the case of the supercrystal, ODLRO can still develop, leading to a
non-vanishing condensate fraction, which we study below.

Translating the classical results for hard spheres~\cite{ReviewZamponi} 
in the quantum case we thus find
that the quantum ground state is unique for $\phi < \phi_K$ and is degenerate
for $\phi > \phi_K$ but the logarithm of the number of ground states is subextensive
and therefore the entropy remains zero above $\phi_K$.  It follows
that the phase transition is not manifested by a non-analyticity in the
free energy of the system.  However, suitable correlation functions 
(such as the ``point to set'' or ``dynamical'' correlation 
functions~\cite{BB04,MS06}) 
should display a growing correlation length at the transition. 
Also the structure function $g(r)$ shows a (weak)
discontinuity at $\phi_K$, but it does not display any long-range 
order.
In the following we will study the properties of this glassy quantum state, 
in particular the ones related to ODLRO. 

\subsubsection{Condensate fraction and corresponding Edwards-Anderson parameter}

Using the results above on the ground-state wavefunction one can
readily obtain the one particle off-diagonal density matrix. As
interference terms can be neglected, it reads:
\begin{equation}
\RR({\xx},{\xx'})=
\sum_\alpha w_\alpha
\;
\RR_\a({\xx},{\xx'})=
V \sum_\alpha w_\alpha
\sum_i n_{i,\a} \;\psi_{i,\a}({\xx})\,\psi^*_{i,\a}({\xx}') 
\;,
\end{equation} 
where the $n_{i,\a}$ and the $\psi_{i,\a}({\xx})$ are, respectively,
eigenvalues and eigenvectors of the one particle density matrix
$\RR_\a(\xx,\xx')$. (Because the wavefunctions $\Psi_\a$ are real, the
$\RR_\a(\xx,\xx')$ are real and symmetric, and the eigenvalues and
eigenvectors are real.) The $n_{i,\a}$ have a physical interpretation:
they are the fraction of particles in the quantum state $i$ within the thermodynamic state $\a$,
and $\sum_{\a,i} w_\a\;n_{i,\a} = 1$ accounts for all the particles in
all possible states $\a$ with probabilities $w_\a$.

For large separations $|\xx-\xx'|\to\infty$, one has for each $\RR_\a$
\begin{equation}\label{rhopsiglass}
\lim_{|{\xx}-{\xx}'|\to\infty} \RR_\a({\xx},{\xx}') 
= f_\a(\rho)\; \frac{\rho_\a({\xx})\,\rho_\a({\xx}') }{\rho^2} = n_{0,\a} \, V
\;  \psi_{0,\a}({\xx}) \, \psi^*_{0,\a}({\xx'})
\;.
\end{equation}

\begin{figure}
\includegraphics[width=8.5cm]{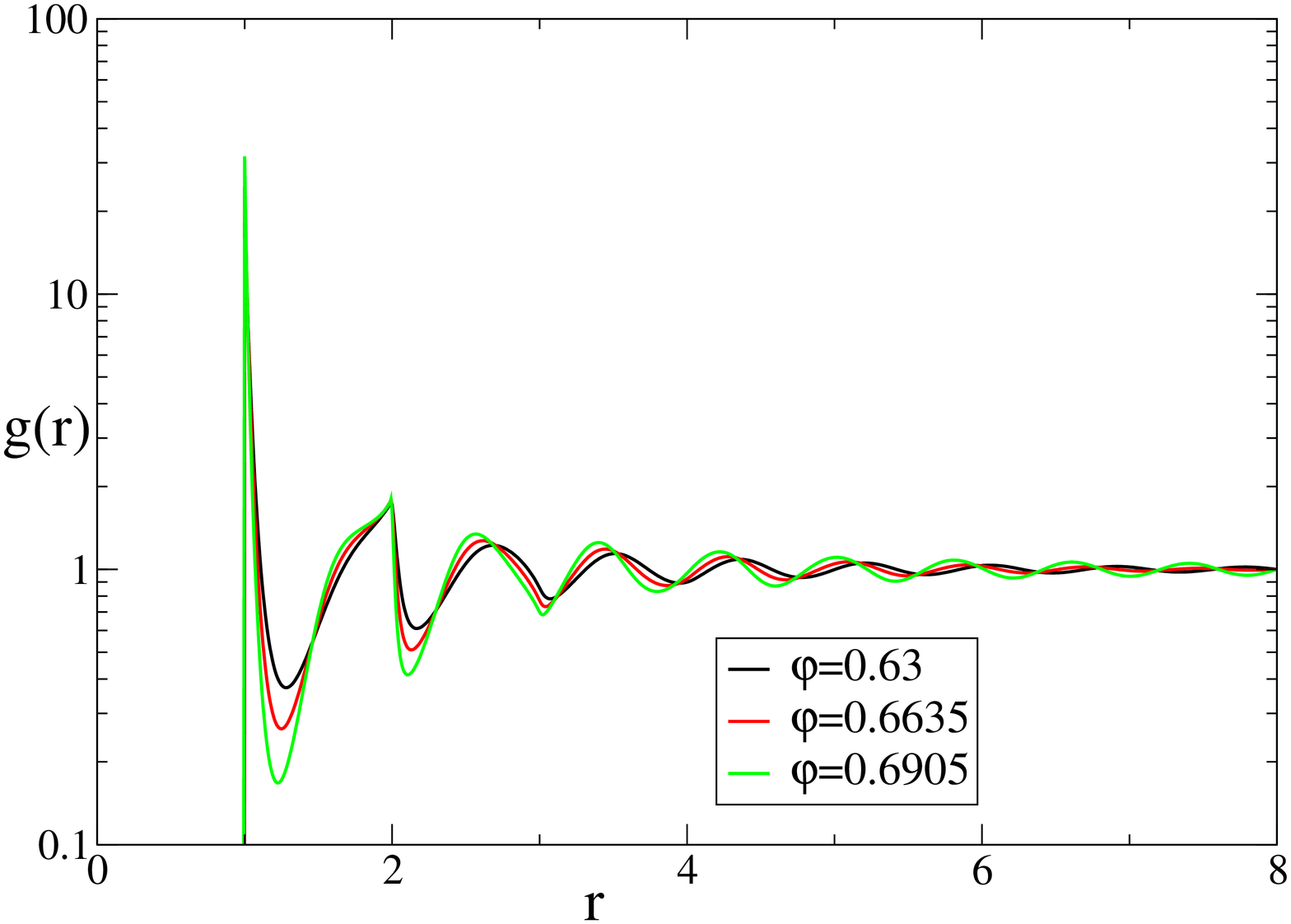}
\includegraphics[width=8.5cm]{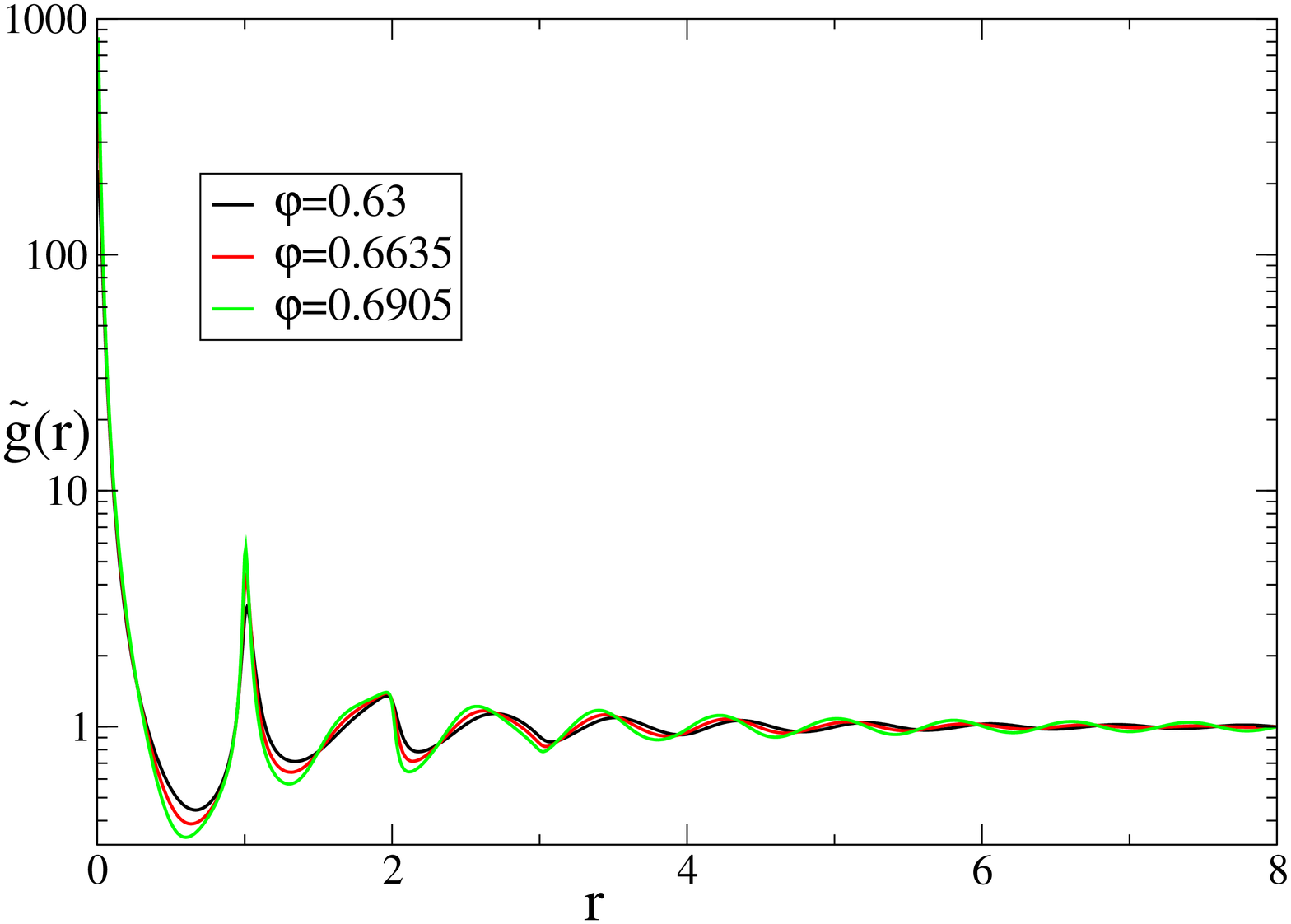}
\caption{Correlation function of the density (left) and of the
condensate wavefunction (right) for the quantum system corresponding
to classical hard spheres at different densities in the glass
phase. Details of the computations are in
Ref.~\onlinecite{ReviewZamponi}.  Note that these quantities can be
measured in Quantum Monte Carlo numerical simulations, see \eg Ref.~\onlinecite{BPS1}.  }
\label{gofr}
\end{figure}

Proceeding similarly to the case of the crystal, we obtain
$\psi_{0,\a}({\xx}) = \sqrt{\frac{f_\a(\r)}{n_{0,\a} V}}
\;\rho_\a({\xx})/\r$. We use the normalization of the eigenvector
$\psi_{0,\a}$ to write
\begin{equation}
 n_{0,\a} = f_\a(\rho) \;\frac{1}{\rho^2} \frac{1}{V}\int d{\xx} 
\;{\rho_\a({\xx})}^2 \ ,
\end{equation}
and for the total condensate fraction we get
\begin{equation}\label{n0glass1}
 n_{0} = \frac{1}{\rho^2} \sum_\a
w_\a\; f_\a(\rho) \;\frac{1}{V}\int d{\xx} 
\;{\rho_\a({\xx})}^2 \ .
\end{equation}

We expect that the $f_\a(\rho)$s are independent of the state
$\alpha$, since they are thermodynamic quantities related to the
addition of one single particle to the system [Recall
Eq.~(\ref{eq:f-def})]. In order to quantify the inhomogeneity of the
glass state one can introduce the correlation function of the density
and condensate wavefunction fluctuations that read respectively:
\begin{equation}\label{defg}
\wt g({\bf x},{\bf x}')  
\equiv \r^{-2} \sum_\a w_\a \r_\a({\bf x})\r_\a({\bf x}') \ , \qquad
G_\psi({\bf x},{\bf x}') 
\equiv \sum_\a w_\a \psi_{0,\a}({\bf x}) \psi_{0,\a}({\bf x}').
\end{equation}
Actually, using eq. (\ref{rhopsiglass}) we find that they are related
one to the other by a simple proportionality relation:
\begin{equation}\label{Gpsi}
\wt g({\bf x},{\bf x}') =\frac{n_0 V}{f(\r)}\sum_\a w_\a \psi_{0,\a}({\bf x}) \psi_{0,\a}({\bf x}') \equiv
\frac{n_0 V}{f(\r)} G_\psi({\bf x},{\bf x}') \ .
\end{equation}
Substituting these definitions in Eq.~(\ref{n0glass1}) we get
\begin{equation}
 n_{0} = f(\r)\frac{1}{V}\int d{\xx} \frac{1}{\rho^2} \sum_\a
w_\a \ \rho_\a(\xx)^2 = f(\r)\frac{1}{V}\int d{\xx} \ \wt g(\xx,\xx) =
f(\r) \wt g(0)  \ ,
\end{equation}
where we used that $\wt g(\xx,\xx') = \wt g(|\xx-\xx'|) = \wt g(r)$,
as translational invariance is restored after averaging over all possible 
states.
Note that $\wt g(\xx,\xx')$ plays the role of
an Edwards-Anderson order parameter for the glass state. 
For instance one can define
\begin{equation}
q_{\rm EA}
=\sum_\a
w_\a\;\frac{1}{V}\int d{\xx} 
\;\left(\frac{\rho_\a({\xx})}{\rho}-1\right)^2 
\; = \wt g(0) - 1 \ .
\end{equation}
Since the density and condensate wavefunction fluctuations are
proportional $\wt g(\xx,\xx')$ also represents (up to a
proportionality constant) the inhomogeneity of the condensate
wavefunction. It is a quantitative measure of how much the condensate
is amorphous. It can be computed in numerical simulation. Look for
example to Fig. 3 of Ref.~\onlinecite{BPS1} which provides a visual
representation of the inhomogeneity captured by $\wt g(\xx,\xx')$.

In order to obtain quantitative results on the condensate fraction we
have to compute $\wt g(0)$ and $f(\r)$. The replica
method~\cite{CFP98b}, in particular the small cage
expansion~\cite{MP99b}, has been successfully applied to describe the
glassy phase of hard spheres~\cite{ReviewZamponi,PZ05}.  These two
quantities can be indeed obtained using the procedure detailed in
Refs.~\onlinecite{ReviewZamponi,PZ05}.  Since this is well
documented~\footnote{There is only one step that is not detailed in
Refs.~\onlinecite{ReviewZamponi,PZ05} so we sketch it here for
completeness, and it is the proof that $\wt g(0) = \frac{1}{\rho}
\frac1{(4\pi A)^{3/2}}$.  In the replica formalism $\wt g(\xx,\xx') =
\r^{-2} \sum_{ij}^{1,N} \la \d(\xx-\xx_i^a) \d(\xx'-\xx_j^b)\ra$,
where $a \neq b$ are two different replicas. In the language of
Refs.~\onlinecite{ReviewZamponi,PZ05}, replicas are arranged in {\it
molecules} labeled by the index $i=1,\cdots,N$. Then if $i\neq j$
particles in different replicas are constrained to be at distance
$|\xx-\xx'| \gtrsim \s$ and do not contribute to $\wt g(\xx,\xx)$.
The only term that contributes is the one for $i=j$, and from Eq.~(4)
in Ref.~\onlinecite{PZ05} we have $\wt g(\xx,\xx') = \r^{-2}
\sum_{i=1}^{N} \la \d(\xx-\xx_i^1) \d(\xx'-\xx_i^2)\ra =\r^{-2} \int
d\xx_3 \cdots d\xx_m \;\r(\xx_1,\cdots,\xx_m)$.  Then the result follows
from Eqs.(11) and (12) of Ref.~\onlinecite{PZ05}.  } we do not
reproduce the computation and just quote the final result which is
very similar to the one for the crystal:
\begin{equation}
\label{n0gla}
n_0= \frac{f(\rho)}{\rho} \frac1{(4\pi A)^{3/2}}
\;,
\end{equation}
where $A$ is the so called cage radius and is a measure of particle
vibrations.  Hence, by using the equation of state for the glass and
the values of $A$ reported in Ref.~\onlinecite{ReviewZamponi,PZ05} we
finally obtain a quantitative result for the glass condensate fraction
which is reported in Fig.~\ref{fig:m}. Note that the results of
Ref.~\onlinecite{ReviewZamponi,PZ05} depend slightly on the
approximation that is used to describe the liquid; for consistency, we
used the results obtained using the Carnahan-Starling approximation as
we did for the liquid.

The replica method allows to compute both $g(r)$ and $\wt g(r)$ in the glass phase, 
but within a different
approximation scheme known as HyperNetted Chain (HNC) 
approximation~\cite{HansenMacDonald}. The results,
taken from Refs.~\cite{CFP98b,ReviewZamponi}, are reproduced in Fig.~\ref{gofr}.

\section{Quantum slow dynamics and the approach to the quantum glass transition}
\label{sec:5}
In the following we shall focus on the real time dynamics of the
superfluid phase when the transition to the superglass is
approached. Also in this case, the knowledge of dynamical correlation
functions for Brownian hard spheres will allow us to obtain results on
dynamical correlation functions of the corresponding quantum problem.
In the next section we will explain how the mapping works for dynamic
observables.
 
\subsection{Mapping from Brownian dynamics of hard spheres 
to real time dynamics of the quantum model}

In order to show how one can obtain information about the real-time
quantum dynamics from the Langevin dynamics (\ref{Langevin}) it is
useful to introduce a bracket notation for the Fokker-Planck
problem; we define $P(\{\xx\},t) = \la \{\xx\} | P(t) \ra$, and we
denote by $|G\ra$ the Gibbs distribution (\ref{GibbsP}), $P_{\rm
G}(\{\xx\},t) =\la \{\xx\} | G \ra = \exp[-U_N(\{\xx\})/T]/Z_N$, such
that $H_{FP} |G\ra =0$. We also denote by $\la + |$ the constant
state, $\la + | \{\xx\} \ra = 1$. Note that
\begin{equation}
H_{FP}^\dag = e^{U_N/T}\; H_{FP}\; e^{-U_N/T} \ ,
\end{equation}
which is consistent with Eq.~(\ref{H-HFP}) and the fact that $H$ is
Hermitian. Then
\begin{equation}
H_{FP} |G\ra = 0
\hskip.5cm \Rightarrow \hskip.5cm
0 = \la G | H_{FP}^\dag = \la G |\, e^{U_N/T} \;H_{FP}\; e^{-U_N/T} 
= Z_N^{-1} \la + | H_{FP}\; e^{-U_N/T}
\hskip.5cm \Rightarrow \hskip.5cm
0 = \la + | H_{FP}
\end{equation}
\ie $\la + |$ is a left eigenvector of $H_{FP}$ with zero eigenvalue. 

One observable that is particularly interesting to characterize the 
quantum dynamics is the dynamical structure factor, $F_{Q}(q,t) $ which is 
the time dependent correlation function of a
Fourier component of the density operator, $\r_q(\{\xx\})=\sum_l e^{i \mathbf {q \cdot x_l}}$. 
We shall show that the dynamical structure factor for the quantum problem is 
related to the imaginary time analytic continuation of the dynamical 
structure factor, $F_{cl}(q,t)$,  for Brownian hard spheres. 

Observing that $\la \{\xx\}| e^{-t H_{FP}} | \{\yy\} \ra$
is the probability of going from $\{\yy\}$ to $\{\xx\}$ in time $t$,
we can write the Brownian correlation function as follows:
\begin{equation}\begin{split}
F_{cl}(q,t) & = \la \r_q(t) \:\r_{-q}(0) \ra \equiv 
\int d\{\xx\}\, d\{\yy\}
\; \r_q(\{\xx\})\; \la \{\xx\}| e^{-t H_{FP}} | \{\yy\} \ra 
\; \r_{-q}(\{\yy\}) \;\frac{ e^{-\b U_N(\{\yy\})}}{Z_N} \\
& 
=\la + |\; \r_q \;e^{-t H_{FP}}\; \r_{-q} |G\ra = 
\la + | e^{t H_{FP}} \;\r_q \;e^{-t H_{FP}}\; \r_{-q} |G\ra  =
\la + | e^{-U_N/(2 T)} \;e^{t H}\; \r_q \;e^{-t H} \;e^{U_N/(2 T)}\; \r_{-q} |G \ra \\
&= \la 0 | e^{t H} \;\r_q \;e^{-t H} \;\r_{-q} |0 \ra 
= \sum_n | \la 0 | \r_q | n \ra |^2 \;e^{-t (E_n-E_0)}
=\int_0^\io \frac{d\omega}{2\pi} \;\rho_q(\omega) \;e^{-\omega t} \ , \\
\end{split}\end{equation}
where $|0\ra \equiv \sqrt{Z_N} \;e^{U_N/(2 T)} |G \ra$ is the quantum
ground state (\ref{Jastrow}), $|n \ra$ are the excited states, and
\begin{equation}
\rho_q(\omega) \equiv 2\pi \sum_n | \la 0 | \r_q | n \ra |^2 \;
\delta(E_n - E_0 - \omega)
\end{equation}
is the distribution of classical (inverse) relaxation times for the density
fluctuations.

We are interested in the quantum correlations; the quantum response
function is given by \cite{forster}
\begin{equation}\begin{split}
R_Q(q,t) & = 
i \th(t) \; \la 0 | \big[ e^{i t H} \;\r_q \;e^{-i t H} ,\r_{-q} \big] |0\ra 
= 
i \th(t) \sum_n | \la 0 | \r_q | n \ra |^2 
\left[ e^{-i t (E_n-E_0) } - e^{i t (E_n-E_0) } \right] \\
& = -2 \th(t) \int_0^\io \frac{d\omega}{2\pi} \;\r_q(\omega) 
\,\sin (\omega t) \ .
\end{split}\end{equation}
It follows that the imaginary part of its Fourier transform reads: 
\begin{equation}
R''_Q(q,\omega) =\frac12\, \text{sgn}(\omega) \;\r_q(|\omega|) \ ,
\end{equation}
and, using the quantum (bosonic) fluctuation-dissipation theorem at zero temperature, we
get the quantum correlation function:
\begin{equation}
S_Q(q,\omega) = \frac12 \,\text{sgn}(\omega)\, R''_Q(q,\omega) = \frac12\,
\r_q(|\omega|)
\hskip.5cm \Rightarrow \hskip.5cm
F_Q(q,t) = \int_{-\io}^\io \frac{d\omega}{2\pi} \; S_Q(q,\omega) 
\; e^{-i \omega t} =
\int_0^\io \frac{d\omega}{2\pi}\; \r_q(\omega)\, \cos (\omega t)
\end{equation}

This final simple expression allows us to obtain results on the dynamical structure factor starting from the 
distribution of (inverse) relaxation times. 

\subsection{Results for the time dependent density-density correlator}

\begin{figure}
\includegraphics[width=8.5cm]{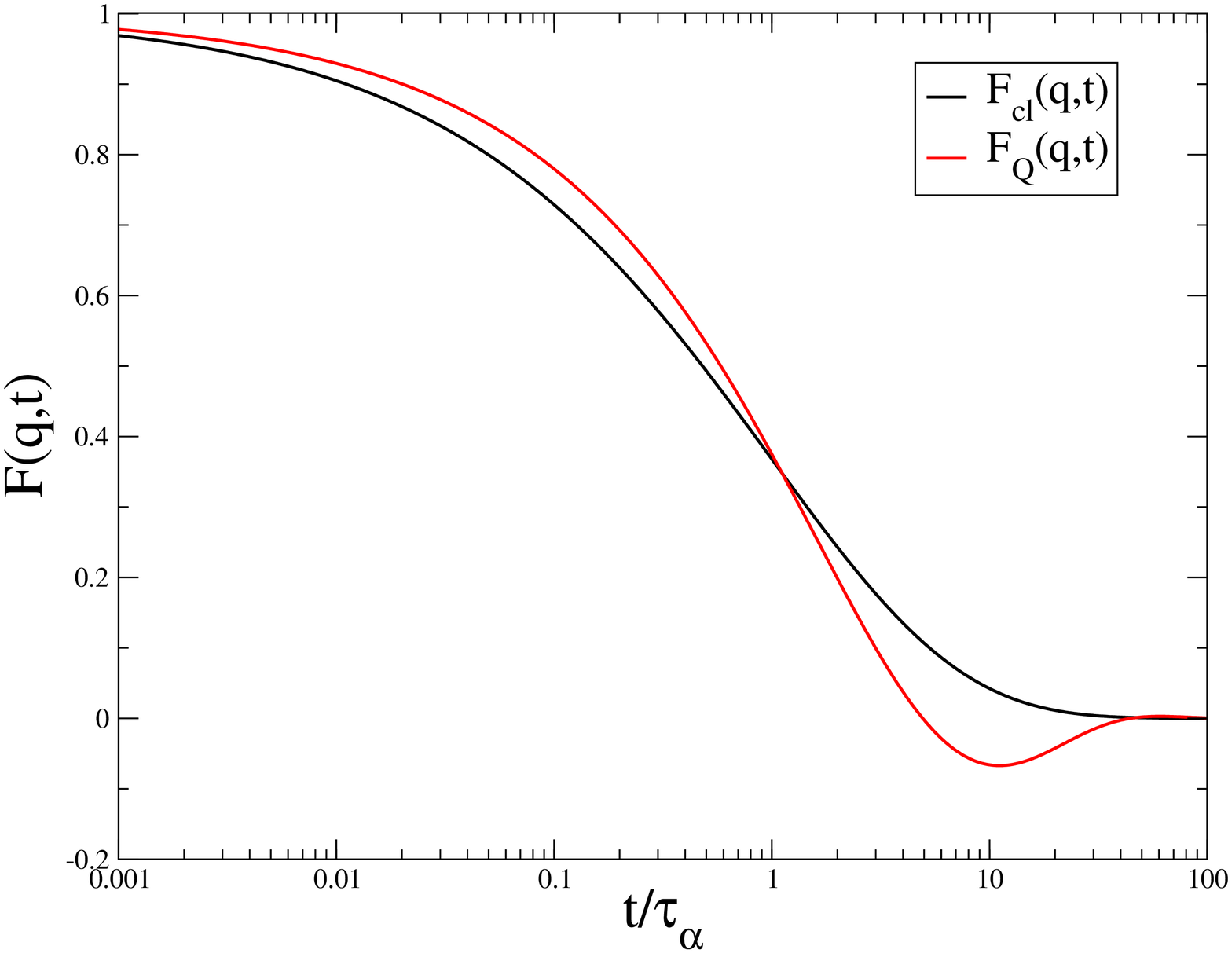}
\includegraphics[width=8.5cm]{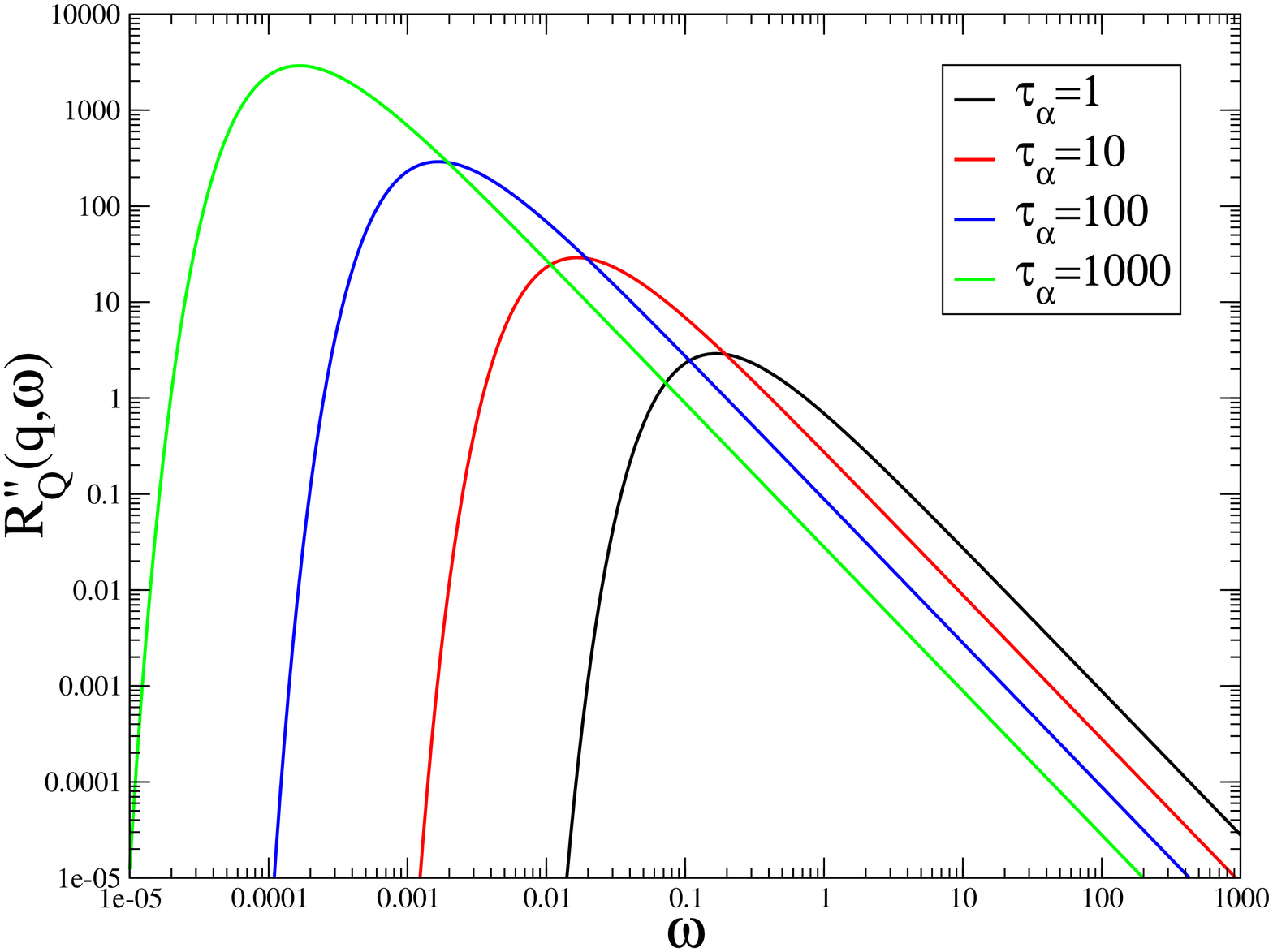}
\caption{ (Left) Long-time shape of the correlation function
Eq.~(\ref{Hstretched}) $F_{cl}(q,t) \sim \exp(-\sqrt{t/\t_\a})$ and
$F_Q(q,t) \sim \exp(-\sqrt{t/2\t_\a}) \, \cos (\sqrt{t/2\t_\a})$ as
a function of $t/\t_\a$ in a linear-log scale. (Right) Imaginary part
of the response function of the quantum problem,
Eqs. (\ref{corr-scaling}) and (\ref{Hstretched}), as a function of
frequency for different values of $\t_\a$ (in arbitrary units).  }
\label{cq}
\end{figure}

Assume that the density correlator has been normalized in such a way
that $F(q,t=0)= 1$; then $\int_0^\io \frac{d\omega}{2\pi}
\rho_q(\omega) = 1$.  In the glass transition literature, different
phenomenological expressions have been used in order to describe the
slowing down of the relaxation of $F(q,t)$ on approaching the glass
transition.

In dielectric spectroscopy one often writes
\begin{equation}
F_{cl}(q,t) = \int_{-\io}^\io d \ln \t \:\; G_q(\ln \t) \; e^{-t/\t} \ ,
\hskip1cm 
G_q(\ln \t) = \frac{\r_q(1/\t)}{2\pi \t} \ ,
\end{equation}
and to describe the slow relaxation (low-frequency) part it is assumed
that $G_q(\ln \t)$ contains a single time scale $\t_\a(q)$,
$G_q(\ln\t) = \left(\t/\t_\a(q)\right)
\;g\left(\t/\t_\a(q)\right)$. The scaling function $g(x)$ might depend
weakly on $q$.  Then
\begin{equation}\label{corr-scaling}
\begin{split}
F_{cl}(q,t) &=  \int_0^\io dx \;g(x)\; e^{-t/(x \t_\a(q))} 
= f_{cl}(t/\t_\a(q)) \ , 
\\
F_Q(q,t) &= \int_{-\io}^\io d\ln \t \:\; G_q(\ln \t) \, \cos(t/\t) = 
\int_0^\io dx \;g(x)\, \cos\left(\frac{t}{x \t_\a(q)}\right) = f_Q(t/\t_\a(q)) \ , \\
R''_Q(q,\omega) &= \frac{\pi}{\omega}\; \frac{1}{\omega \t_\a(q)}\;
g\left(\frac{1}{\omega \t_\a(q)}\right) \ .
\end{split}\end{equation}

Several forms have been used successfully in the literature,
see Ref.~\onlinecite{BLG99} for a review:
\begin{enumerate}
\item {\it Stretched exponential $\b=1/2$} - The simplest non-trivial
case is $f_{cl}(y)=e^{-\sqrt{y}}$; in this case
\begin{equation}\label{Hstretched}
g(x) = \frac{e^{-x/4}}{\sqrt{4\pi x}} \hskip1cm \Rightarrow \hskip1cm
f_Q(y) = e^{-\sqrt{y/2}} \cos \big[\sqrt{y/2}\big] \ .
\end{equation}
\item {\it Stretched exponential} - For $f_{cl}(y)=e^{-y^\b}$ and
generic $\b$ we have
\begin{equation}
g(x) = \frac{1}{\pi x} \int_0^\io ds \; e^{-s-(x s)^\b \cos(\pi \b)} 
\; \sin [ (x s)^\b \sin(\pi \b) ] \ ,
\end{equation}
and in this case $f_Q(y)$ must be computed numerically.
\item {\it Cole-Davidson} - Another common expression, corresponding
to
\begin{equation}
g(x) = \frac{\sin(\pi \g)}{\pi} \left(\frac{x}{1-x}\right)^\g \,
\hskip.5cm 0\leq x\leq 1 \ .
\end{equation}
In this case $f_Q(y)$ can be computed in terms of Hypergeometric functions.
\end{enumerate}

Here we limit ourselves to illustrate the qualitative behavior of
$F_Q(q,t)$ on approaching the glass transition. For this, we consider
solely the first case. (We checked numerically that the other forms
give qualitatively similar results -- oscillations in $F_Q$ are much
more pronounced if one uses the Cole-Davidson form due to the
frequency cutoff.) Some examples constructed using
Eq.~(\ref{Hstretched}) are reported in figure~\ref{cq}. The classical
correlation function decays from 1 to 0 over the time scale $\t_\a$;
the corresponding (real-time) quantum correlation function, directly
accessible in numerical simulations, displays the same slow decay
modulated by oscillations according to (\ref{Hstretched}). The
imaginary part of the response function, directly accessible in
experiments, shows a peak at $\omega \sim 1/\t_\a$, whose amplitude
increases $\propto \t_\a$ upon increasing $\t_\a$.  Note that usually
in classical glassy systems these low-frequency features are
accompanied by faster relaxations, related to ``intra-cage'' motion
and/or fast molecular relaxations. These have been neglected here and
in figure~\ref{cq} but we expect them to be present in the quantum
case as well. They would appear as secondary peaks at higher
frequency, and typically their density dependence is weak. The
complete time-dependence of the density density correlation function
would therefore be a first rapid relaxation to a plateau value and
then a second one which is the one studied in detail in this
section. At the glass transition this second relaxation does not take
place anymore since the relaxation time diverges (or it is larger than
any experimental timescale). As a consequence part of the density
fluctuations becomes frozen. The plateau value in the correlation
function measure precisely that. Actually, the plateau value for a
wave-vector $q$ will coincide with the Fourier transform of the
function $\wt g({\bf x}-{\bf x}') $ defined in Eq.~(\ref{defg}).

\subsection{Condensate Fluctuations}

To conclude this section we will focus on the dynamical fluctuations of
the condensate wavefunction. As we already discussed, in the glass
phase the system can be in many different states, each one
characterized by a different density profile and by a condensate
wavefunction, $\psi_{0,\a}(\xx)$, that in our model is simply
proportional to the classical density profile of the corresponding
state. In Eq.~(\ref{Gpsi}) we defined the correlation function of the
fluctuations of the condensate wavefunction when the system is frozen in a given amorphous state.  
This has the form reported in figure~\ref{gofr} and can be 
accessed by a direct computation of the condensate wavefunction as
done in Ref.~\onlinecite{BPS1}. It is important to notice that 
it can also be obtained from purely dynamical measurements as we will now show.

Consider the time-dependent one particle density matrix; in second
quantization it is defined as $\RR_Q(t; \xx,\xx') \equiv \la
\hat\psi^\dag(\xx,t) \;\hat\psi(\xx',0) \ra$, where $\hat\psi(\xx,t)$
is the standard bosonic annihilation operator. Within first
quantization, it can be written as follows:
\begin{equation}
\label{Rtdef}
\RR_Q(t; \xx,\xx') = V \int d\xx_2 \cdots d\xx_N \: d\xx'_2 \cdots
d\xx'_N \:
\Psi_G(\xx,\xx_2,\cdots,\xx_N)\, 
\la \xx_2,\cdots,\xx_N | e^{-it H} | \xx'_2,\cdots,\xx'_N \ra 
\,\Psi_G(\xx',\xx'_2,\cdots,\xx'_N) \ .
\end{equation} 
Clearly for $t=0$ Eq.~(\ref{Rtdef}) gives back Eq.~(\ref{Rdef}).  In order to understand the time dependent behavior 
it is useful to consider the evolution in imaginary time and use again the mapping
on the Langevin dynamics. From (\ref{H-HFP}) we have
\begin{equation}
\label{ruota} 
\la \xx_2,\cdots,\xx_N | \, e^{-t H} \, | \xx'_2,\cdots,\xx'_N \ra =
[\Psi_G(\xx_2,\cdots,\xx_N)]^{-1} \la \xx_2,\cdots,\xx_N | 
\, e^{-t H_{FP}} \, | \xx'_2,\cdots,\xx'_N \ra 
\;\Psi_G(\xx'_2,\cdots,\xx'_N) \ ,
\end{equation} 
where
$\Psi_G(\xx_2,\cdots,\xx_N)$ is the Jastrow state (\ref{Jastrow}) for
$N-1$ particles.  Plugging (\ref{ruota}) in (\ref{Rtdef}) and using
the explicit form (\ref{Jastrow}), we finally obtain
\begin{equation}\label{Rlangevin}
\begin{split}
\RR(t; \xx,\xx') &= 
V \int d\xx_2 \cdots d\xx_N \:d\xx'_2 \cdots d\xx'_N \ \
e^{-\frac12 \sum_{i=2}^N V(\xx - \xx_i) } \ \ 
\la \xx_2,\cdots,\xx_N | e^{-t H_{FP}} | \xx'_2,\cdots,\xx'_N \ra \\
& \times e^{-\frac12 \sum_{i=2}^N V(\xx' - \xx'_i) }
\ \ \frac{e^{-U_{N-1}(\xx'_2,\cdots,\xx'_N)}}{Z_N} \ ,
\end{split}\end{equation}
where $U_{N-1}$ is the interaction potential of the $N-1$ particles and we have used the notation $\RR$ for the imaginary time 
continuation of $\RR_Q$.

In the special case of a Jastrow hard sphere wavefunction,
the factors of $\frac12$ are irrelevant. Then Eq.~(\ref{Rlangevin}) becomes
\begin{equation}
\RR(t; \xx,\xx') = 
V \int d\xx_2 \cdots d\xx_N \: d\xx'_2 \cdots d\xx'_N \
e^{- \sum_{i=2}^N V(\xx - \xx_i) } \ 
\la \xx_2,\cdots,\xx_N | e^{-t H_{FP}} | \xx'_2,\cdots,\xx'_N \ra 
\;\frac{e^{-U_N(\xx',\xx'_2,\cdots,\xx'_N)}}{Z_N} \ ,
\end{equation}
and has a straightforward interpretation in terms of Brownian dynamics
of hard spheres: one should pick up a configuration $\{ \xx' \}$ from
the equilibrium distribution, such that particle 1 is in $\xx'_1 =
\xx'$. Then particle 1 must be removed, and particles $2,\cdots,N$
evolved according to the Langevin dynamics in absence of particle
1. Finally, one should attempt to reintroduce particle $1$ at position
$\xx_1=\xx$. The function $\RR(t; \xx,\xx')$ is the probability that
the attempt is successful, or in other words that there is a void at
time $t$ around $\xx$ large enough to allow the reinsertion of
particle 1.

Close to the glass transition, as we already discussed, there is a
huge separation of time scales in the classical dynamics between a ``fast''
intra-state relaxation and a ``slow'' relaxation corresponding to
hopping between different states and characterized by a growing time
scale $\t_\a$. If we remove particle $1$ at $t=0$, the perturbation of
the density field will not relax inside the initial state until $t \sim
\t_\a$, where the state will change.  This implies that $\RR(t;
\xx,\xx')$ will have a {\it plateau} at times $\t_{\rm fast} \ll t
\lesssim \t_\a$ corresponding to the stationary probability {\it
inside the initial state at $t=0$}. This is given by
\begin{equation}
\RR(\t_{\rm fast} \ll t \lesssim \t_\a; \xx,\xx') \propto
\sum_{\a} w_\a \;\psi_{0,\a}(\xx)\; \psi_{0,\a}(\xx') \propto \wt g(\xx-\xx') \ ,
\end{equation}
where one has to average over all possible initial states.  As
obtained in the previous section for the density-density correlation
function, we expect that in presence of a huge separation of
timescales, as it is the case close to the glass transition (or in the
glass state where the second relaxation does not take place anymore),
one finds that $\RR(\t_{\rm fast} \ll t \lesssim \t_\a; \xx,\xx')\sim
\RR_Q(\t_{\rm fast} \ll t \lesssim \t_\a; \xx,\xx')$. One way to
understand this result consists in expressing $\RR(t)$ in terms of the
Fourier transform of $\RR_Q$: $\RR(t)=\int \frac{d\omega}{2\pi}\;
\RR_Q(\omega)\exp(-\omega t)$. If $\RR$ displays a very long plateau
this means that $\RR_Q(\omega)$ contains two distinct contributions
corresponding to $\omega\propto 1/\t_{\rm fast}$ and $\omega\propto
1/\tau_\alpha$. As a consequence, $\RR_Q$ on times intermediate
between fast and slow timescales will coincide with $\RR$ since the
contribution from large frequencies (of the order or $1/\t_{\rm
fast}$) will have died out and the contribution from the very long
frequencies (of the order of $1/\tau_{\alpha}$) will be the same since
$\exp(-i\omega t)\simeq\exp(-\omega t)\simeq 1$. As a conclusion,
$\RR_Q$ will display a plateau whose extension will become infinite
beyond the glass transition. The value of $\RR_Q$ on the plateau
corresponds to the fraction of frozen condensate wavefunction
fluctuations and equals $G_\psi$ (and hence is proportional to $\tilde
g$). Therefore this quantity can also be computed in a dynamic
framework without introducing replicas.

\section{Superfluid properties and phase diagram of realistic superglass phases}
\label{sec:6}

The conclusion from the previous sections is that the ground state of
the model can be a liquid, a crystal or a glass and all these phases
are characterized by a finite condensate fraction $n_0$.  However, the
study of the superfluid properties requires also the knowledge of
excited states, or at least of the excitation spectrum. The latter is
related to superfluid properties by the celebrated Landau
argument~\cite{LandauIX} that predicts for the critical velocity
\begin{equation}\label{landau}
v_c \leq \min_k [ \ee(k)/k ] \ .
\end{equation}
In the zero-temperature liquid He$^4$ phase, the excitation spectrum
is linear at small $k$ and has a minimum at larger $k$, therefore
$v_c$ is finite and the system is superfluid.

Unfortunately, for the Jastrow wavefunction discussed above,
Eq.(\ref{Jastrow}), one can show that the excitation spectrum is
quadratic at small $k$, at least if the potential $V(\xx)$ has finite
integral. Therefore Eq.(\ref{landau}) implies that $v_c=0$ and the
system is not superfluid, much as it happens for an ideal Bose gas
despite a condensate fraction equal to 1 at zero temperature.

The quadratic spectrum of Jastrow wavefunctions can be related to the
following properties of the Hamiltonian (\ref{quantumH}):
\begin{enumerate}
\item The ground state energy per particle $e(\r)$ of the Jastrow
ground state is always zero, therefore the pressure $P=\r^2
\frac{de}{d\r} =0$ and the compressibility is infinite, $\chi_T^{-1} =
\r \frac{dP}{d\r} = 0$.
\item Consequently the sound velocity $c = 1/\sqrt{\r \chi_T} = 0$,
\ie there are no phonons and the linear part of the spectrum at small
$k$ is absent; this is because no restoring force for density
fluctuations is present if $e(\r)$ is independent of $\r$.
\item The static structure factor of the Jastrow wavefunction has the
property $\lim_{k\to 0} S(k) \neq 0$; therefore, the Feynman relation~\cite{Huang}
$\ee(k) = k^2/(2 m S(k))$ gives $\ee(k) \propto k^2$, consistent
with $c=0$.
\item Finally, it is possible to identify a hidden symmetry in the
problem, related to the special form of the potential in
(\ref{quantumH}), that is responsible for non-trivial cancellations in
the Bogoliubov low-density perturbation theory for (\ref{quantumH}) around the
ideal gas limit. Again these cancellations are responsible for the
absence of the linear part of the spectrum.
\end{enumerate}

When using Jastrow wavefunctions as variational functions for liquid
Helium, a classical strategy \cite{FCR70} to reintroduce phonons is to add a
non-integrable part to the Jastrow potential, such that $V(\xx) \sim
|\xx|^{-2}$ at large $|\xx|$. In this way $S(k) \sim k$ at small $k$
and the excitation spectrum is linear. This is also quite natural since
long range correlations are expected in the ground state of generic quantum
systems even if the original interaction is short ranged.
The inclusion of these terms in our formalism is possible and their presence 
does not influence
much the results for the properties of the glass, \eg for 
$g(r)$ and $\wt g(r)$ discussed in section 
\ref{sec:glass-phase}, except at large $r$/small $k$.

However, the classical-quantum mapping 
will give in this case a quantum Hamiltonian (\ref{quantumH})
with very long ranged interaction, while we would like to
keep the original local nature of the quantum Hamiltonian.
\begin{figure}
\includegraphics[width=8.5cm]{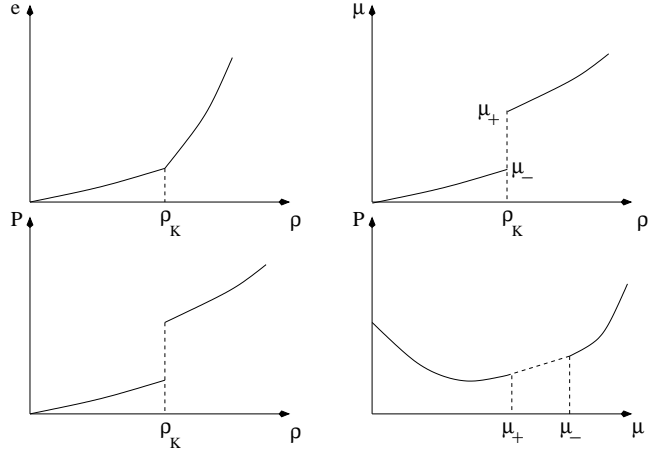}
\caption{Schematic behavior of $e(\r)$, $P(\mu)$, $\mu(\r)$ and
$P(\r)$ across the glass transition.  Note that there are values of
$P$ and $\mu$ that correspond to the same density $\r_K$.  }
\label{P}
\end{figure}
Therefore we consider an alternative way
to solve the problem: we propose to introduce a perturbation
of the quantum Hamiltonian (\ref{quantumH}) by adding a small potential term 
$\D \VV_N(\{\xx\}) = \sum_{i<j} \d u(\xx_i-\xx_j)$ and treat it in perturbation
theory.

The situation here is quite similar to Bogoliubov low-density perturbation theory\cite{AGD}:
indeed, the ideal Bose gas has a finite condensate fraction (actually equal to 1) 
but is not superfluid because its excitation spectrum is quadratic. Once an infinitesimal 
interaction is added (or in the very dilute regime), the spectrum immediately becomes
linear and the system becomes superfluid\cite{AGD}.
Hence by analogy we argue that perturbation theory can be applied in our case.
As a check of this argument we verified
that in the low-density limit, as discussed above, the Bogoliubov theory
of the unperturbed model leads to a quadratic spectrum due to a hidden symmetry.
On the contrary, the perturbation $\D \VV_N$ breaks the hidden
symmetry, and in its presence the usual Bogoliubov
theory applies and leads to a linear spectrum.

Using perturbation theory
at first order, it is straightforward to show that 
\begin{equation}
e(\rho) = \frac{\r}2 \int dr\; g(r) \;\d u(r) \ ,
\end{equation}
$g(r)$ is the correlation function of the hard sphere liquid at
density $\r$. One can show by explicit computation for a suitable
specific form of $\d u(r)$ that, at the glass transition density
$\r_K$, the ground state energy is continuous but its first derivative
has a jump.

The pressure $P(\mu)$ (as a function of the chemical potential $\mu$)
is the Legendre transform of $e(\r)$:
\begin{equation}
P(\mu) = \max_\r [ \r \m - \r e(\r) ] \ ,
\end{equation}
therefore
\begin{equation}\begin{split}
\mu(\rho) &= \frac{d}{d\r} [ \r e(\r) ] \ , \\
P(\rho) &= \r^2 \frac{d}{d\r} e(\rho) \ . \\
\end{split}\end{equation}
From the above expressions one can see that $P(\mu)$ is continuous and
convex as $e(\r)$, while $\mu(\r)$ and $P(\r)$ have a jump in $\r_K$,
see Fig.~\ref{P}. Therefore the glass transition, that is a second
order transition in the classical case, looks like a first order
transition in the zero-temperature quantum problem. This is similar to
previous results on quantum mean field glass models with quenched
disorder, see Refs. \onlinecite{CGS01,BC01}.  On the other hand the
properties of the first order transition are quite different.  
In the case of mean field quantum glass models the glass and
liquid are really different phases and the glass phase does not appear
(via density fluctuations that are frozen on timescales diverging at
the transition) in a continuous way from the liquid, contrary to what
happens for the superglass.

The sound velocity is now determined by
\begin{equation}
c^2 = \frac1{\r \chi_T} = \frac{d P}{d \r} = \frac{d}{d\r} \left[ \r^2 \frac{de}{d\r} \right] \neq 0 \ .
\end{equation}
General arguments~\cite{Huang} show that if $c\neq 0$, then $S(k) \leq
k/(2 m c)$; assuming equality the Feynman formula gives $\ee(k) = c
k$, consistently with the existence of sound waves.

Therefore the system in this case has a finite critical velocity and
is superfluid.  Note that in the glass the first peak of the structure
factor, that determines $\min_k [ \ee(k)/k ]$, is close to the one of
the liquid, suggesting that the critical velocity should stay close to
the one of the liquid. Still it is known (\eg in Helium) that the precise
determination of the critical velocity is complicated and depends on the 
geometry, therefore estimating 
precisely its value in the superglass phase is beyond the scope of this paper.

\section{Conclusion}

In this paper we presented and analyzed a concrete model of
interacting bosons that displays unambiguously, in addition to superfluid and
supercrystal phases, metastable superfluid and superglass
phases. This shows concretely that a system can be at the same time
glassy, {\it i.e.} displaying very slow dynamics for the structural
degrees of freedom, and supersolid, {\it i.e.} showing a superfluid
component. Note that there is no paradox, exactly as there is none for
the supercrystal.  Indeed, Leggett in his original paper\cite{Leggett}
``Can a solid be superfluid?'' explicitly mentioned the possibility of
super-amorphous solids.

The exact ground state of the model we considered is known by
construction: it is a Jastrow wavefunction with a hard sphere potential
form. We constructed the quantum Hamiltonian for which this
wavefunction is exact via a mapping to Brownian motion of classical
hard spheres, following the connection between the Fokker-Planck
operator for the classical stochastic dynamics and the Schr\"odinger
operator describing the associated quantum system. This mapping allows
us to understand the physics of the superglass phase by using results
from the well-studied problem of densely-packed hard spheres.

Our findings, which were summarized in Sec.~\ref{Summary}, are most
simply conveyed through Fig.~\ref{fig:phasediagram}. By changing the
density of particles, one goes from a superfluid phase to a
supercrystal via a first order phase transition. If the density is
increased fast or in case of (binary) mixtures, one reaches a
metastable superfluid state, and if the density is further increased,
one reaches the superglass phase.

The analysis we carried out for the hard sphere Jastrow wavefunction
can be extended for other systems for which the Jastrow wavefunction
corresponds to a classical potential leading to glassy
dynamics. Classically, mixtures may become glasses at certain
compositions, for example 80\%/20\% mixtures of two species of
particles interacting via Lennard-Jones
potentials~\cite{KobAndersen}. The problem of binary mixtures is
interesting in that one could potentially realize the system in cold
atomic gases, where the relative fractions of two species could be
controlled. We have worked out the correspondence between the
classical and quantum problems for binary mixtures in
Appendix~\ref{sec:appendix}. Following the same reasoning as in our
work on the hard sphere wavefunctions, the binary mixtures can display
superglass behavior as long as the mapped classical problem does so.

The usefulness of the classical to quantum mapping is that it allows
us to make precise and concrete statements about the nature of the
ground state of the devised Hamiltonian. By construction, the ground
state energy is exactly zero, and thus one must add a small
perturbation so as to obtain a non-zero speed of sound, as we have
done in Sec.~\ref{sec:6}. We have added a 2-body potential as
perturbation, because we could make use of the knowledge of the
density-density correlation $g(r)$ in obtaining the effect of the
additional interaction on the ground state energy, pressure,
compressibility, and sound velocity of the system. It is an
interesting possibility that if, as opposed to the 2-body
perturbation, one adds a 3-body potential, one could compensate the
3-body term that arises from the classical to quantum mapping. In
other words, we speculate that by removing (or reducing) the 3-body
term by a compensating perturbation, one could achieve a quantum
Hamiltonian with a glassy ground state, only 2-body terms, and a
finite sound velocity.  This could perhaps be tested numerically, by
considering the system with the potential in Eq.~(\ref{pairpotential})
alone. Moreover, if the short-ranged ``sticky'' part of the
Eq.~(\ref{pairpotential}) potential is removed, this may not
substantially change the results. At least, we expect the system to be
glassy as the original one (the question of superfluidity is instead a
more tricky one and needs further investigations).  The intuitive
reason is as follows: for dense packings and a hard sphere potential,
if one turns down the kinetic energy (say by ``increasing'' the mass
of the particles), thus making the system more classical, one should
recover the physics of the classical hard sphere system. Because the
packings are dense, turning up the kinetic energy (say by
``decreasing'' the mass of the particles), should not favor very much
transitions between a glassy state $\a$ to another $\b$. Because of
the hard sphere potential (basically infinite on the scales of the
kinetic energy), we do not expect, at a dense packing, that quantum
fluctuations will reduce considerably (at least for not too small $m$)
the timescales to escape from classical glassy configurations.

Concerning the superglass transition, we found it to be first order
but quite unusual.  It is different from the quantum first order
transitions found for mean field glassy models in Refs.
\onlinecite{CGS01,BC01}. It would be interesting to develop and study
mean field models (or mean field approximations) able to reproduce our
results and the superglass phase transitions. One of the main motivations is that
they could allow one to analyze the finite temperature regime which is
clearly out of reach of the approach developed in this work.  In
particular, it would be very interesting to apply Quantum Mode
Coupling Theory\cite{ReichmanRabani} (QMCT) to the model we focus
on. It would be very interesting to know whether QMCT predicts quantum
correlation functions that are connected to the one predicted by
classical MCT via the classical quantum-mapping we employed.  If yes,
this would mean that QMCT is able to capture this unusual first order
phase transition toward the superglass phase and it would allow one to
study its finite temperature extension.

Finally, let us discuss preliminary results for He$^4$ which is one of
the original motivations of our work.  Jastrow wavefunctions have been
used extensively to describe the ground state of He$^4$ in both
fluid~\cite{McMillan,FCR70} and solid~\cite{HL68} phases. Therefore it
is tempting to try to study the superglass phase of He$^4$ by using
the Jastrow wavefunction as a variational wavefunction. For He$^4$ it
was found that the liquid phase is well described by a potential $V(r)
\sim r^{-5}$. We then computed the glass transition density for this
potential following Ref.~\onlinecite{CT05}.  We found that the superglass 
transition takes place at a density $\r_K \sim 0.45
{\AA}^{-3}$ and a very small jump in the first derivative of $e(\r)$
at $\r_K$, the relative variation of $e'(\r)$ being of the order of
$10^{-3}$. Unfortunately, the value of $\r_K$ is 10 times higher than
the density at which the superglass has been observed in
Ref.~\onlinecite{BPS1}. However, it is well known that the Jastrow
wavefunction overestimates the liquid-crystal transition for He$^4$
and, more generally, provides a poor description of the solid phases.
A more refined investigation should involve Shadow-like
wavefunctions~\cite{Shadow}, and we plan to report on such a study in
a future publication where more details on the Jastrow wavefunction
will also be given.

\acknowledgments We thank S. Balibar, S.~Baroni, M.~Boninsegni, 
J.-P. Bouchaud, G.~Carleo, C.~Castelnovo,
A. Lef\`evre, S.~Moroni, O. Parcollet, L.~Reatto, D.~R.~Reichman, 
M.~Tarzia, and X. Wayntal for
discussions. We also thank R.~Di Leonardo for bringing Ref.~\onlinecite{BLG99} to our 
attention.
GB work is supported by ANR grant DYNHET. FZ acknowledges financial support from the
ESF Research Networking Programme INSTANS.

\appendix
\section{Quantum-classical mapping for a binary mixture of bosons}
\label{sec:appendix}

Here we work out the details of the connection between the quantum and
classical fluids for a mixture of two types of particles, $A$ and
$B$. We start again with the Jastrow type wavefunction written in
terms of the interaction among $N=N_A+N_B$ particles, through a
potential
\begin{equation}
U(\{\xx\};\{{\mathbf X}\})\equiv
U({\mathbf x}_1,\dots,{\mathbf x}_{N_A};{\mathbf X}_1,\dots,{\mathbf X}_{N_B})
=
\sum_{i>j} V_{AA}(|{\mathbf x}_i-{\mathbf x}_j|)
+
\sum_{I>J} V_{BB}(|{\mathbf X}_I-{\mathbf X}_J|)
+
\sum_{i,I} V_{AB}(|{\mathbf x}_i-{\mathbf X}_J|)
\; ,
\end{equation}
where $i,j=1,\dots,N_A$ and $I,J=1,\dots,N_B$. The dynamics of a
classical mixture of particles interacting through this potential is
clearly a particular case of the general Langevin equation specified
by Eqs.~(\ref{Langevin}) and (\ref{pot_gen}), where we set $\g_i =
\g_A$ and $\g_I = \g_B$, and the form of the Fokker-Planck operator
follows straightforwardly from Eq.~(\ref{HFP}).

The quantum Hamiltonian is, from (\ref{quantumH}),
\begin{eqnarray}
H&=&
\sum_i \frac{p_i^2}{2m_A} 
+
\sum_I \frac{P_I^2}{2m_B} 
+
{\cal V}(\{\xx\};\{{\mathbf X}\})
\;,
\end{eqnarray}
where $m_{A,B} = \hbar^2 \g_{A,B}/(2 T)$ and, setting $T=1$
and $\hbar =1$,
\begin{equation}
{\cal V}(\{\xx\};\{{\mathbf X}\})\equiv
{\cal V}({\mathbf x}_1,\dots,{\mathbf x}_{N_A};{\mathbf X}_1,\dots,{\mathbf X}_{N_B})=
\frac{1}{4m_A} \sum_i
\left\{
 \frac{1}{2} \left(\nabla_i U \right)^2
- \nabla^2_i U
\right\}
+
\frac{1}{4m_B} \sum_I
\left\{
\frac{1}{2}  \left(\nabla_I U \right)^2
-  \nabla^2_I U
\right\}
.
\end{equation}
The ground state state of this Hamiltonian is
\begin{equation}
\Psi_G(\{\xx\};\{{\mathbf X}\})
=
\frac{1}{\sqrt{{Z_{N_A,N_B}}}}
\;\exp\left[-\frac{1}{4}\;
U(\{\xx\};\{{\mathbf X}\})
\right]
\;.
\end{equation}
The quantum potential ${\cal V}(\{\xx\};\{{\mathbf X}\})$ will again
have contributions in the form of 2-body and 3-body terms:
\begin{eqnarray}
\VV&=&
\sum_{i>j} v^{AA\,\rm pair}_{ij}
+
\sum_{I>J} v^{BB\,\rm pair}_{IJ}
+
\sum_{i,I} v^{AB\,\rm pair}_{iI}
+
\text{3-body terms}
\;.
\end{eqnarray}
The pair potential is a simple generalization of Eq.~(\ref{eq:pair}):
\begin{subequations}
\begin{eqnarray}
v^{AA\,\rm pair}_{ij}
&=& 
\frac{1}{2m_A}\left\{
-\frac{d-1}{r_{ij}}\,V_{AA}'(r_{ij})
-V_{AA}''(r_{ij})
+\frac{1}{2}\, [V_{AA}'(r_{ij})]^2
\right\}
\\
v^{BB\,\rm pair}_{IJ}
&=& 
\frac{1}{2m_B}\left\{
-\frac{d-1}{r_{IJ}}\,V_{BB}'(r_{IJ})
-V_{BB}''(r_{IJ})
+\frac{1}{2}\, [V_{BB}'(r_{IJ})]^2
\right\}
\\
v^{AB\,\rm pair}_{iI}
&=& \frac{m_A+m_B}{4m_Am_B}\left\{
-\frac{d-1}{r_{iI}}\,V_{AB}'(r_{iI})
-V_{AB}''(r_{iI})
+\frac{1}{2}\, [V_{AB}'(r_{iI})]^2
\right\}
\;,
\end{eqnarray}
\end{subequations}
and similarly for the 3-body terms.

\subsection{Quantum model associated with Lennard-Jones binary mixtures}

\begin{figure}
\includegraphics[width=8.5cm]{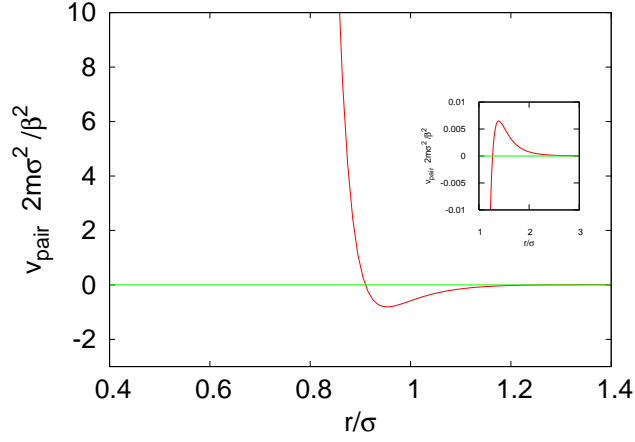}
\caption{Form of the pair potential $v^{\rm pair}(r)$ in the mapped
quantum problem that derives from the classical $\a=6$ and
$\b=12$ Lennard-Jones potential. ($\e=1$ is set for simplicity.) The
potential is repulsive at large and small distances, but to see the
repulsion for large $r$ one needs to zoom closer, as shown in the
inset.}
\label{fig:pair-potential-LJ}
\end{figure}

Let us construct the quantum model associated to a Lennard-Jones
binary mixture, for which the underlying classical system can be
glassy, for instance certain 80\%/20\% mixtures of $A$/$B$ particles
\cite{KobAndersen}.
Let us consider classical potentials of the form
$V_{P_1P_2}(r)=-\e_{P_1P_2}\left[(r/\s_{P_1P_2})^{-\a}-(r/\s_{P_1P_2})^{-\b}\right]$,
where $P_1,P_2=A$ or $B$, with three energy scales, $\e_{AA},\e_{BB}$
and $\e_{AB}$, and three characteristic lengths, $\s_{AA},\s_{BB}$ and
$\s_{AB}$, in the problem. The exponents $\b>\a>0$ make the classical
potential attractive at long distances and repulsive at short
ones. The corresponding quantum pair potential is
\begin{eqnarray}
v^{P_1P_2\,\rm pair}(r)
= \frac{m_{P_1}+m_{P_2}}{4m_{P_1}m_{P_2}}\;
\frac{\e_{P_1P_2}}{\s_{P_1P_2}^2}
\left\{
(\a(\a+2-d)\,(r/\s_{P_1P_2})^{-(\a+2)}
-
(\b(\b+2-d)\,(r/\s_{P_1P_2})^{-(\b+2)}
\right.
\nonumber\\
\left.
+
\e_{P_1P_2}
\left[
\frac{\a^2}{2}\,(r/\s_{P_1P_2})^{-(2\a+2)}
-
\a\b\,(r/\s_{P_1P_2})^{-(\a+\b+2)}
+
\frac{\b^2}{2}\,(r/\s_{P_1P_2})^{-(2\b+2)}
\right]
\right\}
\;.
\end{eqnarray}
Notice that the pair potential is repulsive for both small and large
distances: $v^{P_1P_2\,\rm pair}(r)\sim (r/\s_{P_1P_2})^{-(\a+2)}$ for
$r/\s_{P_1P_2}\gg 1$ and $v^{P_1P_2\,\rm pair}(r)\sim
(r/\s_{P_1P_2})^{-(2\b+2)}$ for $r/\s_{P_1P_2}\ll 1$. At intermediate
distances, $r/\s_{P_1P_2}\sim 1$, the potential can become attractive,
as illustrated in Fig.~\ref{fig:pair-potential-LJ} starting from an
$\a=6$ and $\b=12$ classical Lennard-Jones potential.

\subsection{Off-diagonal long range order in binary mixtures}
\label{appendix:binary-ODLRO}

We can extend the Penrose-Onsager definition of off-diagonal long
range order~\cite{Onsager} to a mixture of two distinct types of bosonic
atoms. Let
\begin{subequations}
\begin{equation}
\RR_A({\mathbf x}-{\mathbf x}')
=
V\int
\prod_{i=2}^{N_A} d{\mathbf x}_i
\;\prod_{I=1}^{N_B} d{\mathbf X}_I
\;
\Psi_G({\mathbf x},
{\mathbf x}_2,\dots,{\mathbf x}_{N_A};{\mathbf X}_1,\dots,{\mathbf X}_{N_B})
\,
\Psi_G({\mathbf x}',
{\mathbf x}_2,\dots,{\mathbf x}_{N_A};{\mathbf X}_1,\dots,{\mathbf X}_{N_B})
\end{equation}
and
\begin{equation}
\RR_B({\mathbf X}-{\mathbf X}')
=
V\int
\prod_{i=1}^{N_A} d{\mathbf x}_i
\;\prod_{I=2}^{N_B} d{\mathbf X}_I
\;
\Psi_G({\mathbf x}_1,\dots,{\mathbf x}_{N_A};
{\mathbf X},
{\mathbf X}_2,\dots,{\mathbf X}_{N_B})
\,
\Psi_G({\mathbf x}_1,\dots,{\mathbf x}_{N_A};
{\mathbf X}',
{\mathbf X}_2,\dots,{\mathbf X}_{N_B})
\;.
\end{equation}
\end{subequations}

We can write these expressions in terms of the classical liquid
correlations:
\begin{eqnarray}
\RR_A({\mathbf x}-{\mathbf x}')
=
&&
V\,\frac{1}{Z_{N_A,N_B}}
\int
\prod_{i=2}^{N_A} d{\mathbf x}_i
\;\prod_{I=1}^{N_B} d{\mathbf X}_I
\;
e^{-U({\mathbf x}_2,\dots,{\mathbf x}_{N_A};{\mathbf X}_1,\dots,{\mathbf X}_{N_B})}
\nonumber\\
&&
\;\;\;\;\;\;\;\;\;\;\;\;
\times
e^{-\frac{1}{2}\left[
\sum_{j=2}^{N_A}  V_{AA}(|{\mathbf x}-{\mathbf x}_j|)
+
\sum_{J=1}^{N_B}  V_{AB}(|{\mathbf x}-{\mathbf X}_J|)
\right]}
\;
e^{-\frac{1}{2}\left[
\sum_{j=2}^{N_A}  V_{AA}(|{\mathbf x}'-{\mathbf x}_j|)
+
\sum_{J=1}^{N_B}  V_{AB}(|{\mathbf x}'-{\mathbf X}_J|)
\right]}
\\
=
&&
V\,\frac{Z_{N_A-1,N_B}}{Z_{N_A,N_B}}
\int
\prod_{i=2}^{N_A} d{\mathbf x}_i
\;\prod_{I=1}^{N_B} d{\mathbf X}_I
\;
P_G({\mathbf x}_2,\dots,{\mathbf x}_{N_A};{\mathbf X}_1,\dots,{\mathbf X}_{N_B})
\nonumber\\
&&
\;\;\;\;\;\;\;\;\;\;\;\;
\times
e^{-\frac{1}{2}\left[
\sum_{j=2}^{N_A}  V_{AA}(|{\mathbf x}-{\mathbf x}_j|)
+
\sum_{J=1}^{N_B}  V_{AB}(|{\mathbf x}-{\mathbf X}_J|)
\right]}
\;
e^{-\frac{1}{2}\left[
\sum_{j=2}^{N_A}  V_{AA}(|{\mathbf x}'-{\mathbf x}_j|)
+
\sum_{J=1}^{N_B}  V_{AB}(|{\mathbf x}'-{\mathbf X}_J|)
\right]}
\\
=
&&
V\,\frac{Z_{N_A-1,N_B}}{Z_{N_A,N_B}}
\;\;\langle
e^{-\frac{1}{2} \Phi_A({\mathbf x})}
\;
e^{-\frac{1}{2} \Phi_A({\mathbf x}')}
\rangle
\end{eqnarray}
where we defined the potential caused by the $N_A-1$ type $A$ particles
and the $N_B$ type $B$ ones as
\begin{equation}
\Phi_A({\mathbf x})
=
\int \!\! d{\mathbf r}\;
\left[
V_{AA}(|{\mathbf x}-{\mathbf r}|)\,\rho_A({\mathbf r})
+
V_{AB}(|{\mathbf x}-{\mathbf r}|)\,\rho_B({\mathbf r})
\right]
\;,
\end{equation}
with the densities given by $\rho_A({\mathbf r})=\sum_{j=2}^{N_A}
\delta({\mathbf r}-{\mathbf x}_j)$ and $\rho_B({\mathbf
r})=\sum_{J=1}^{N_B} \delta({\mathbf r}-{\mathbf x}_j)$.

Similarly, one obtains
\begin{equation}
\RR_B({\mathbf X}-{\mathbf X}')
=
V\,\frac{Z_{N_A,N_B-1}}{Z_{N_A,N_B}}
\;\;\langle
e^{-\frac{1}{2} \Phi_B({\mathbf X})}
\;
e^{-\frac{1}{2} \Phi_B({\mathbf X}')}
\rangle
\end{equation}
where the potential caused by the $N_A$ type $A$ particles
and the $N_B-1$ type $B$ is given by
\begin{equation}
\Phi_B({\mathbf X})
=
\int \!\! d{\mathbf r}\;
\left[
V_{AB}(|{\mathbf X}-{\mathbf r}|)\,\rho_A({\mathbf r})
+
V_{BB}(|{\mathbf X}-{\mathbf r}|)\,\rho_B({\mathbf r})
\right]
\;.
\end{equation}

One can define the condensate fraction of type $A$ bosons as
$n_A=\RR_A(\infty)\equiv \lim_{|{\mathbf x}-{\mathbf x}'|\to\infty}
\RR_A({\mathbf x}-{\mathbf x}')$, and so
\begin{equation}
n_A=V\,\frac{Z_{N_A-1,N_B}}{Z_{N_A,N_B}}
\;\;\langle
e^{-\frac{1}{2} \Phi_A({\mathbf 0})}
\rangle^2
\;,
\end{equation}
where we assumed that the two-point correlation function factorizes,
\begin{equation}
\langle e^{-\frac{1}{2} \Phi_A({\mathbf x})} \; e^{-\frac{1}{2}
\Phi_A({\mathbf x}')} \rangle\to \langle e^{-\frac{1}{2}
\Phi_A({\mathbf x})}\rangle \; \langle e^{-\frac{1}{2} \Phi_A({\mathbf
x}')} \rangle = \langle e^{-\frac{1}{2} \Phi_A({\mathbf 0})}
\rangle^2
\;,
\end{equation}
and also that the one-point function is translational
invariant. Notice that the factorization assumption should be fine for
short-ranged potentials or potentials that decay sufficiently fast,
but there may be certain (possibly pathological) potentials for which
it may fail.

One can check that this definition of the condensate fraction of $A$
bosons is such that $n_A\le 1$ as follows. First we use H{\"o}lder's
inequality to obtain
\begin{equation}
\langle e^{-\frac{1}{2} \Phi_A({\mathbf x})} \; e^{-\frac{1}{2}
\Phi_A({\mathbf x}')} \rangle
\le
\sqrt{
\langle e^{-
\Phi_A({\mathbf x})}\rangle
\langle e^{-
\Phi_A({\mathbf
x}')} \rangle
}
= \langle e^{-\Phi_A({\mathbf 0})}
\rangle
\;,
\end{equation}
from which we conclude that $\RR_A({\mathbf x}-{\mathbf x}')\le
\RR_A({\mathbf 0})=1$, the last equality following trivially from the
normalized wavefunctions (plus translational invariance). Thus,
$n_A\le 1$ follows from the definitions above. Similar results apply
to $n_B$.

Notice that $\RR_A({\mathbf 0})=1$ is a simple way to obtain $\langle
e^{-\Phi_A({\mathbf 0})} \rangle=
\frac{1}{V}\,\frac{Z_{N_A,N_B}}{Z_{N_A-1,N_B}}$; we will use this equality below.

\subsubsection{Fixing the chemical potential}

If we work at fixed chemical potential, we can relate the condensate
fraction of bosons $A$ and $B$ to the fugacities $z_{A,B}$ and the
densities $\rho_{A,B}$ as follows. The grand canonical partition
function for the binary mixture is given by
\begin{equation}
{\cal Z}=\sum_{N_A,N_B} \frac{{z_A}^{N_A}}{N_A!}\frac{{z_B}^{N_B}}{N_B!}
\;Z_{N_A,N_B}
\;.
\end{equation}
At the saddle point, with particle numbers $N_A^*$ and $N_B^*$, one
has the relations
\begin{equation}
\frac{N^*_A}{z_A}\;Z_{N^*_A-1,N^*_B}
=Z_{N^*_A,N^*_B}
=\frac{N^*_B}{z_B}\;Z_{N^*_A,N^*_B-1}
\;.
\label{eq:fugacity-rel}
\end{equation}
Substitution of these relations in the equations for $n_{A,B}$
allows us to write
\begin{subequations}
\begin{equation}
n_A=
\frac{z_A}{\rho_A}
\;\;\langle
e^{-\frac{1}{2} \Phi_A({\mathbf 0})}
\rangle^2
\end{equation}
and
\begin{equation}
n_B=
\frac{z_B}{\rho_B}
\;\;\langle
e^{-\frac{1}{2} \Phi_B({\mathbf 0})}
\rangle^2
\;.
\end{equation}
\end{subequations}
Notice that Eq.~(\ref{eq:fugacity-rel}) also allows us to write
\begin{equation}
\langle e^{-\Phi_A({\mathbf 0})} \rangle=
\frac{1}{V}\,\frac{Z_{N_A,N_B}}{Z_{N_A-1,N_B}}=\frac{\rho_A}{z_A}
\qquad
{\rm and}
\qquad
\langle e^{-\Phi_B({\mathbf 0})} \rangle=
\frac{1}{V}\,\frac{Z_{N_A,N_B}}{Z_{N_A,N_B-1}}=\frac{\rho_B}{z_B}
\end{equation}
at equilibrium.

These relations allows us further to put lower bounds on the
condensate fraction. To do so, consider without loss of generality
potentials such that $\Phi_{A,B}({\mathbf 0})\ge 0$; one can always do
so by shifting the energies by a constant value such that the
potentials $V_{AA},V_{BB}$ and $V_{AB}$ are non-negative (and using
that the densities $\rho_{A,B}\ge 0$). In this case, we can write
\begin{equation}
\langle
e^{-\frac{1}{2} \Phi_{A,B}({\mathbf 0})}
\rangle
\ge
\langle
e^{- \Phi_{A,B}({\mathbf 0})}
\rangle
=
\frac{\rho_{A,B}}{z_{A,B}}
\end{equation}
and thus
\begin{equation}
n_{A,B}\ge
\frac{\rho_{A,B}}{z_{A,B}}
\;.
\label{eq:bound}
\end{equation}



For a hard sphere potentials, the equality $ \langle e^{-\frac{1}{2}
\Phi^{\rm hard}_{A,B}({\mathbf 0})} \rangle = \langle e^{- \Phi^{\rm
hard}_{A,B}({\mathbf 0})} \rangle$ is satisfied. Thus, hard spheres
have the lowest possible condensate fractions respecting the bound
Eq.~(\ref{eq:bound}).

\end{document}